\documentclass[namedreferences]{SolarPhysics}

\usepackage[optionalrh,natbib]{spr-sola-addons}
\usepackage{epsfig}     
\usepackage{graphicx}
\usepackage{graphicx}
\usepackage{color}                       
\usepackage{url}    
\usepackage{a4}    

\usepackage{solaheader}	
\newcommand{\solareceiveddate}{28 Jul 2011} 
\newcommand{\solaaccepteddate}{11 Oct 2011}

\renewcommand{\vec}{\mathbf}
\newcommand{\eq}[1]{(\ref{#1})}

\begin{document}

\begin{article}

\begin{opening}
\received{A} 
\accepted{1}

\title{The Free Energy of NOAA Solar Active Region AR 11029}
\author{S.A. ~\surname{Gilchrist}\sep
        M.S. ~\surname{Wheatland}\sep
        K.D.~\surname{Leka}      
        }
\institute{S.A. Gilchrist \sep M.S. Wheatland \\ 
Sydney Institute for Astronomy, School of Physics, The University of Sydney, NSW 2006, Australia \\
e-mail: s.gilchrist@physics.usyd.edu.au\\}

\institute{K.D. Leka \\ 
NorthWest Research Associates, CoRA Division, 3380 Mitchell Lane, Boulder, CO 80301, USA }
            
\begin{abstract}

\noindent The NOAA active region AR 11029 was a small but highly active 
sunspot region which produced 73 {\it GOES} soft X-ray flares during its transit of the
disk in late October 2009. The flares appear to show a departure from the 
well known power-law frequency-size distribution. Specifically, too few 
{\it GOES} C-class and no M-class flares were observed by comparison with a power-law 
distribution (Wheatland in {\it Astrophys. J.} {\bf 710}, 1324, {2010}). 
This was conjectured to be due to the region having insufficient magnetic 
energy to power the missing large events. We construct nonlinear force-free extrapolations
of the coronal magnetic field of active region AR 11029
using data taken on 24 October by the SOLIS Vector-SpectroMagnetograph (SOLIS/VSM),
and data taken on 27 October by the {\it Hinode} Solar Optical Telescope 
SpectroPolarimeter ({\it Hinode}/SP). Force-free modeling with 
photospheric magnetogram data encounters problems because the magnetogram 
data are inconsistent with a force-free model. We employ a recently developed 
`self-consistency' procedure which addresses this problem and 
accommodates uncertainties in the boundary data (Wheatland and R{\'e}gnier in 
{\it Astrophys. J.} {\bf 700}, L88, {2009}). We calculate the total energy and free energy 
of the self-consistent solution which provides a model for the coronal 
magnetic field of the active region. The free energy of the region is found to be 
$\approx 4 \times 10^{29}\,{\rm erg}$ on 24 October, and $\approx 7\times 10^{31}\,{\rm erg}$
on 27 October. An order of magnitude scaling between RHESSI non-thermal energy 
and {\it GOES} peak X-ray flux is established from a sample of flares from
the literature and is used to estimate flare energies from observed 
{\it GOES} peak X-ray flux. Based on the scaling, we conclude that the estimated 
free energy of AR 11029 on 27 October when the flaring rate peaked 
is sufficient to power M-class or X-class flares, and hence the modeling 
does not appear to support the hypothesis that the absence of large flares 
is due to the region having limited energy.
\end{abstract}


\end{opening}


%
%
\section{Introduction}
\label{intro}

The dynamics of the Sun's diffuse outer atmosphere, the corona, is
dominated by the solar magnetic field \citep{2004psci.book.....A}. 
Solar flares are the result of the explosive release of magnetic energy 
from intense coronal magnetic fields associated with sunspots. 
Understanding the storage and release of magnetic energy in the corona is fundamental 
to understanding solar flares, and other coronal transient events ({\it e.g.} 
coronal mass ejections and transient X-ray brightenings). 

\begin{figure}
\centerline{\includegraphics[scale =1.]{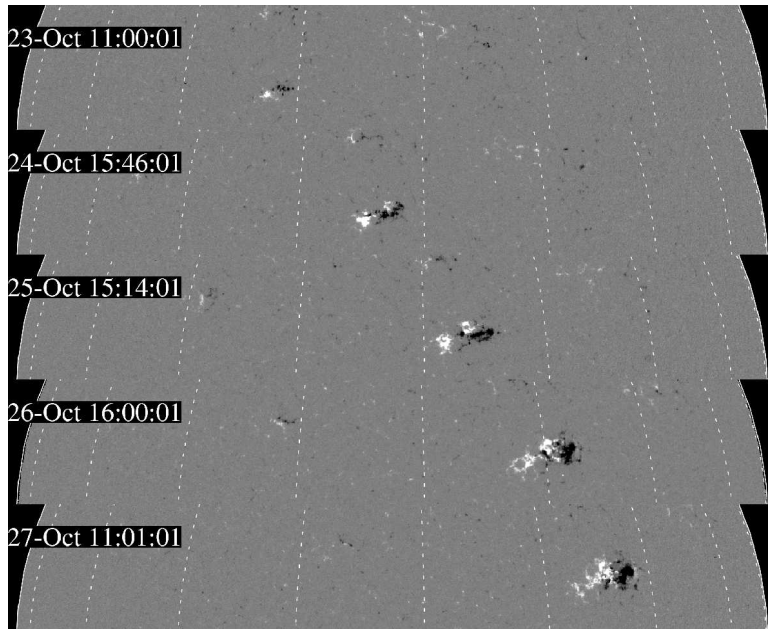}}
\caption{Evolution of active region AR 11029 from 23 to 27 October 2009. The figure shows a 
collage of magnetograms from the  Michelson Doppler Interferometer (MDI) 
aboard the {\it Solar and Heliospheric Observatory}. The data show 
the line-of-sight magnetic field $\approx 120\,{\rm km}$ above the base of the photosphere, and
the field values saturate at $\pm 500\,{\rm gauss \,(G)}$. The times when the data are 
taken are indicated in the panels.} 
\label{f0}
\end{figure}

The NOAA solar active region AR 11029 emerged on 21-22 October 2009, 
began flaring on 24 October 2009, and rotated off the
disk on 1 November 2009. Figure \ref{f0}  shows the evolution of 
the photospheric magnetic field of AR 11029 between 23 and 27 October  
as observed by the Michelson Doppler Interferometer (MDI) aboard the 
{\it Solar and Heliospheric Observatory}. Over the period shown the region 
grew in size and complexity. A detailed summary of the region's evolution during 
its transit of the disk is given by \citet{2010ApJ...710.1324W}.  

Active region AR 11029 was highly active, producing a large number of small flares during 
a period of pronounced solar minimum \citep{2009EOSTr..90..257L}. In 
total 73 soft X-ray events are listed 
in the event catalog\footnote{The {\it GOES} event catalog 
is found online at \url{http://www.ngdc.noaa.gov}. } compiled from
observations by the {\it Geostationary Operational 
Environmental Satellites} ({\it GOES}) during the region's transit of the disk. 
The majority are small {\it GOES} B-class events (peak 1-8 $\, {\rm \AA}$ 
soft X-ray flux in the range $10^{-7}{\rm Wm^{-2}}$ to 
$10^{-6}{\rm Wm^{-2}}$). The largest is a {\it GOES} C2.2 event (peak soft 
X-ray flux of $2.2 \times 10^{-6} {\rm Wm^{-2}}$). No moderate sized 
M-class flares (peak flux $10^{-5}\,{\rm Wm^{-2}}$ to $10^{-4}\,
{\rm Wm^{-2}}$) or large X-class flares (peak flux 
$\ge 10^{-4}\,{\rm Wm^{-2}}$) were produced.

The frequency and size of solar flares (by `size' we
mean a measure of the magnitude {\it e.g.} {\it GOES} peak flux) are related by a well 
known power-law distribution ({\it e.g.} \citealt{1956PASJ....8..173A,1991SoPh..133..357H}). 
This distribution is universal in the sense 
that the same power law index is observed at all 
times, for events occurring over the entire Sun and for events that 
occur in different active regions \citep{2000ApJ...532.1209W}. 
The power law in size reflects an underlying power law in the distribution of flare
energies \citep{1991SoPh..133..357H}.  

Analysis of the statistics of the flares produced by active region 
AR 11029 using the {\it GOES} event data provides strong evidence for a 
departure from the expected power-law frequency-peak flux distribution 
for this region \citep{2010ApJ...710.1324W}. The region produced fewer 
large events than expected given the number of small events, based on the 
power law distribution, and hence did not appear to follow the universal 
distribution. 

\begin{figure}
\centerline{\includegraphics[scale =.85]{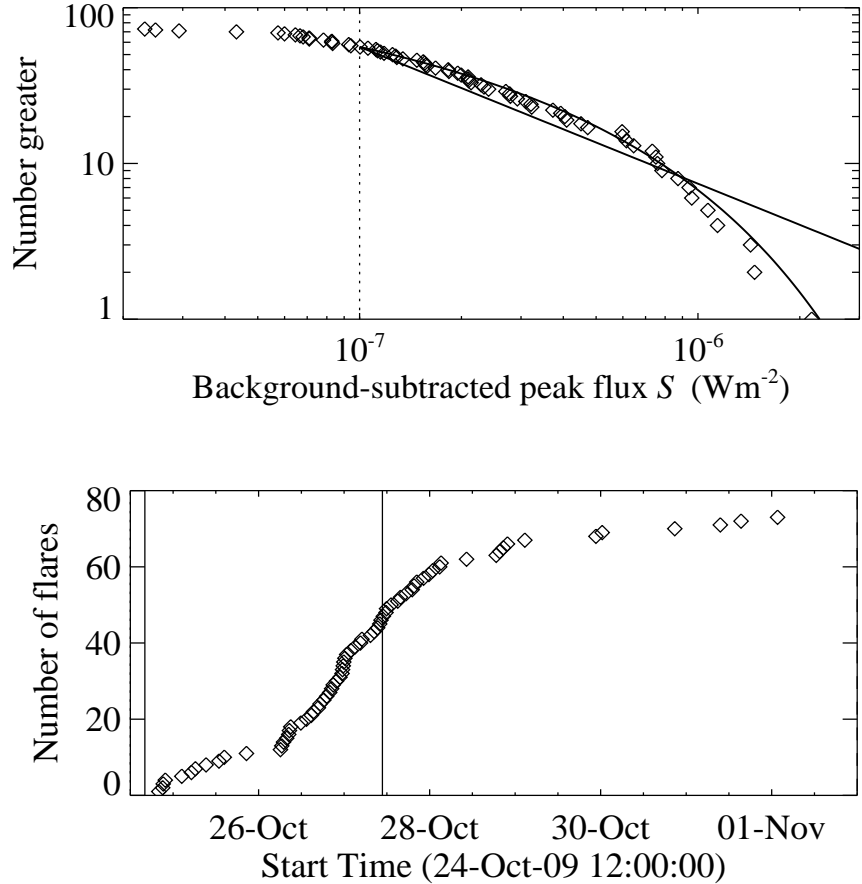}}
\caption{The upper panel shows the frequency-size distribution of soft X-ray flares for
active region AR 11029 \citep{2010ApJ...710.1324W}, plotted as a cumulative number
distribution for the background subtracted peak soft X-ray flux of  
{\it GOES} events. The straight line shows a simple power law distribution, 
and the curved line shows a power law with an exponential rollover at 
large peak fluxes. The dotted vertical line shows the size threshold
above which the models are assumed to apply ($10^{-7}\,{\rm Wm^{-2}}$). 
A Bayesian model comparison favors the rollover model to
the straight power law with an odds ratio of $\approx220$ \citep{2010ApJ...710.1324W}. 
The lower panel shows the cumulative number of {\it GOES} flares observed in 
active region AR 11029 versus time, between 24 October when the 
region began flaring, and 1 November when it rotated off the disk. The times when 
SOLIS/VSM and {\it {\it Hinode}}/SP magnetogram data are available are 
indicated in the lower panel by the solid vertical lines.} 
\label{f1}
\end{figure}

The upper panel of Figure \ref{f1} shows the frequency-peak flux distribution for the 
flares produced by AR 11029 as a cumulative histogram (diamonds), and 
the curves show two model distributions. The straight line corresponds to 
a power-law model distribution $P(S)\sim S^{-\gamma}$ where $S$ is the peak flux
and $\gamma$ is the power-law index. The curved line corresponds to a power 
law model with an exponential rollover at large event sizes
$P(S) \sim S^{-\lambda}\exp(-S/\sigma)$ where $\lambda$ is the power-law
index for small $S$ and $\sigma$ is the rollover size. The 
vertical dotted line shows the peak flux above which the two models are 
assumed to apply ($10^{-7}\,{\rm Wm^{-2}}$). There is a noticeable 
departure from the power law at the large end of the scale. 
Given the data, a Bayesian model comparison taking into account all events showed 
that the rollover model was favored over the simple power-law model 
with an odds ratio\footnote{The odds 
ratio is the relative probability that the data favors the 
rollover model over the power law model: a value of one indicates 
both models are equally likely, and a value of $\approx 220$ indicates 
that the rollover model is $\approx 220$ times more likely given the data 
\citep{2003prth.book.....J}.} of $\approx 220$ \citep{2010ApJ...710.1324W}. 

The absence of large flares was attributed by \citet{2010ApJ...710.1324W} 
to the region having insufficient energy to produce large events. 
In principle an upper bound on flare size must exist which reflects the 
finite energy stored in the active region's magnetic field 
\citep{1991SoPh..133..357H}. However, it has proven difficult to 
observationally identify this effect in flare statistics 
\citep{2007ApJ...663L..45H}. 

By determining the energy stored in the coronal magnetic field of active 
region AR 11029 it is in principle possible to test the hypothesis that 
the absence of large flares results from a lack of energy stored in the 
active region's magnetic field. The energy available for release in flares
is called the free energy $E_{\rm f}$, and is stored in the component of
the coronal magnetic field generated by large scale electric 
currents in the corona \citep{2002A&ARv..10..313P}.
Equivalently, the free energy is the energy in excess of the 
current free (potential) component {\it i.e.} 

\begin{equation}
E_{\rm f} = E - E_0 = \frac{1}{2\mu_0} \int |\mathbf B|^2 dV - \frac{1}{2\mu_0} \int |\mathbf B_0|^2 dV ,
\label{energyFree}
\end{equation}
  
\noindent where $E$ is the total energy of the active region magnetic field $\mathbf B$, 
$E_0$ is the energy of the current free (potential) field $\vec B_0$, and the
integrals are over the coronal volume of the active region. 
The potential field is uniquely defined in a finite volume by the 
normal component of the magnetic field on the closed surface
surrounding this volume. 

Determining the energy of an active region from Equation 
(\ref{energyFree}) requires knowledge of the magnetic field over the 
three dimensional coronal volume of the region. This information 
is difficult to obtain observationally. Solar magnetic fields affect 
the polarization of the Sun's optical emissions through the Zeeman 
effect \citep{2004ASSL..307.....L}. This provides one diagnostic tool 
for inferring the solar magnetic field from spectropolarimetric 
measurements of magnetically sensitive lines, if such lines are 
available. Vector magnetograms are two-dimensional maps of the 
vector magnetic field $\vec B$ close to the level of the photosphere,
derived from spectropolarimetric measurements \citep{2004ASSL..307.....L}. 
The field values are not coronal, but photospheric values. The coronal magnetic field cannot be 
determined observationally except in special cases ({\it e.g.}\ \citealt{1964ApJ...140..817H}), 
although a number of new methods based on radio 
({\it e.g.} \citealt{1997SoPh..174...31W,1998ApJ...501..853L}) 
and infrared ({\it e.g.} \citealt{2004ApJ...613L.177L}) observations are under 
development. 

The lack of observational methods for determining the coronal 
magnetic field motivates modeling of the coronal field from the 
photospheric vector magnetogram data [see reviews by \citet{1989SSRv...51...11S} or more 
recently \citet{2008JGRA..11303S02W}, for an overview of the so called 
`reconstruction problem'], and such models permit estimates of the free energy of
the coronal magnetic field ({\it e.g.} \citealt{2007ApJ...669L..53R,2008A&A...488L..71T}). 
The magnetic field on the lower boundary of the coronal volume is provided by the 
spectropolarimetric observations and this formally defines a boundary value 
problem for determining the coronal field, given a suitable model 
\citep{2006A&A...446..691A}. In this way the three dimensional 
coronal magnetic field can be `extrapolated' from two dimensional 
magnetogram data.   

Nonlinear force-free (NLFF) models of the coronal magnetic field assume 
the field is static and force free ({\it e.g.} \citealt{1994ppit.book.....S}). 
This means all non-magnetic forces are neglected and the magnetic 
Lorentz force is assumed to be self balancing, that is 

\begin{equation}
\mathbf J \times \mathbf B = 0,
\end{equation}

\noindent where $\mathbf J$ is the coronal electric current density. 
In practice a numerical approach is used to solve the force-free 
equations. A number of numerical methods have been developed 
\citep{2008JGRA..11303S02W} and tested on known solutions 
\citep{2006SoPh..235..161S}. Section \ref{intro_forcefree} provides 
further details of force-free modeling.  

Force-free modeling encounters a number of problems in application to 
magnetogram data \citep{2009ApJ...696.1780D}.  This appears to be 
primarily a result of the inconsistency between the force-free models 
and the boundary data. Whilst the corona may be close to force-free \citep{2001SoPh..203...71G}, 
the photosphere is subject to significant non-magnetic forces \citep{1995ApJ...439..474M,2001SoPh..203...71G}. 
\citet{1995ApJ...439..474M} estimated that the solar atmosphere does 
not become force free until at least $400\,{\rm km}$ above the base of the photosphere. 
\citet{2009ApJ...700L..88W} developed a procedure based on an 
implementation of the Grad-Rubin method for calculating nonlinear 
force-free fields \citep{2007SoPh..245..251W} which addresses the 
inconsistency between the data and the model. This approach identifies 
a force-free magnetic field solution with boundary conditions close to, 
but not exactly matching, the magnetogram boundary conditions. The 
resulting `self-consistent' field may represent the coronal magnetic 
field. The self-consistency procedure can accommodate  
uncertainties in the boundary data (details of this method are given in 
Section \ref{intro_sc}). This method has previously been applied 
to vector magnetogram data without incorporating the 
uncertainties in the data \citep{2009ApJ...700L..88W}, and has recently 
been applied with uncertainties \citep{2011ApJ...728..112W}. 

In this paper we use the \citet{2009ApJ...700L..88W} self-consistency 
procedure to reconstruct the coronal magnetic field of AR 11029 from 
vector magnetogram data. The magnetograms are derived from 
spectropolarimetric data provided by the US National Solar Observatory's 
(NSO) Synoptic Optical Long term Investigations of the Sun Vector-SpectroMagnetograph (SOLIS/VSM), 
and the Solar Optical Telescope SpectroPolarimeter onboard the {\it Hinode} 
spacecraft ({\it Hinode}/SP). Based on the reconstructed coronal magnetic 
field we calculate the total energy stored in the 
active region and the free energy $E_{\rm f}$, defined by Equation \eq{energyFree}, 
and compare these results to a lower bound on the energy corresponding to the 
size of the largest observed {\it GOES} event produced by AR 11029. 

This paper is structured as follows. Section 2 provides background on 
nonlinear force-free models and the self-consistency method, and 
Section 3 explains the use of the SOLIS/VSM and {\it Hinode}/SP data. Section 4 
explains the application of the force-free model to the SOLIS/VSM and 
{\it Hinode}/SP data and presents the results of the modeling.
Section 5 discusses the results, and Section 6 presents the conclusions. 
The Appendix contains a glossary of term used throughout the paper. \\

%
%
\section{Background}
\label{background}

\subsection{Force-free Modeling of the Coronal Magnetic Field}
\label{intro_forcefree}

\noindent The difficulties associated with observational determination 
of the coronal magnetic field have motivated numerical extrapolation 
methods which aim to `reconstruct' the coronal magnetic field
from photospheric spectropolarimetric data. This requires a model of 
the coronal magnetic field. 

The nonlinear force-free model is a static model which assumes non-magnetic 
forces in the corona ({\it e.g.} gravity and pressure forces) are negligible compared to the 
magnetic (Lorentz) force. The latter assumption is approximately met when the plasma 
beta, defined as $\beta = 2\mu_0 p/|\mathbf B|^2$, where $p$ is the gas pressure, 
is much less than unity. A force-free field may be  
defined as the solution\footnote{In this paper when we refer to a `model' we mean the nonlinear force-free model defined
by these equations. By `solution' we mean a solution to these equations, for specific boundary
conditions. The Appendix provides a glossary of terms.} to the equations

\begin{equation}   
\mathbf B \cdot \nabla \alpha = 0
\label{ff_lor}
\end{equation}

\noindent and

\begin{equation}
\nabla \times \mathbf B = \alpha \mathbf B
\label{ff_ampere}
\end{equation}

\noindent \citep{1994ppit.book.....S}. Here $\mathbf B$ is the magnetic field vector and 
$\alpha$ is the force-free parameter, which is related to the electric current 
density $\mathbf J = \mu_0^{-1} \nabla \times \mathbf B$ by 
the relationship $\mathbf J = \alpha \mathbf B /\mu_0$. 

Equations (\ref{ff_lor}) and (\ref{ff_ampere}) are nonlinear in the 
general case (when $\alpha$ is a function of position, {\it i.e.} $\alpha=\alpha(\mathbf r)$ 
where $\mathbf r$ is the position vector). 
The nonlinearity, and the mixed elliptic-hyperbolic character of the equations
makes them difficult to solve, motivating a numerical approach. 
Several numerical methods for solving Equations (\ref{ff_lor}) and (\ref{ff_ampere}) 
in a solar context have been developed [see reviews by \citet{1989SSRv...51...11S} 
and \citet{2008JGRA..11303S02W}]. A version of the current-field iteration or 
Grad-Rubin approach, originally proposed by \citet{gr}, is used here 
\citep{2007SoPh..245..251W}. 

The boundary conditions for solving the force-free equations 
are described by \citet{bin} and \citet{gr}. In the solar context, the 
boundary conditions are $B_n$, the component of the magnetic field 
normal to the photosphere and $\alpha_0$, the value of $\alpha$ on the 
photosphere over one polarity of the field {\it i.e.} over the region where 
$B_n >0$, or the region where $B_n <0$. We refer to the two choices, defined 
by the two different polarities, as the $P$ and $N$ choices respectively. 

On the scale of an active region the curvature of the photosphere is 
often ignored \citep{1990SoPh..126...21G}, and Equations (\ref{ff_lor}) 
and (\ref{ff_ampere}) are solved in the Cartesian half space $z \ge 0$, where
$z=0$ is the photosphere. In this case $B_n=B_z$, and the force-free
parameter $\alpha$ on the photosphere is defined by 

\begin{equation}
\alpha_0 = \left. \frac{1}{B_z} \left (\frac{\partial B_y}{\partial x} - \frac{\partial B_x}{\partial y} \right ) \right |_{z=0}.
\label{alpha0}
\end{equation}

\noindent The right hand side of Equation (\ref{alpha0}) may be 
estimated from photospheric vector magnetogram data, which we assume provides 
the vector components of $\mathbf B$ at $z = 0$. 

The boundary value problem outlined in this section is for an infinite 
half space. To compute a numerical solution it is necessary to solve 
the equations on a finite grid of points. This introduces the problem 
that boundary conditions on $\mathbf B$ and $\alpha$ are unknown on the 
top and sides of the finite domain and force-free extrapolation codes 
must deal with this problem. The code used here assumes the magnetic 
field is periodic in the $x$ and $y$ directions, and that the force-free 
parameter $\alpha$ is zero on the top and side boundaries \citep{2007SoPh..245..251W}. 
Equation (\ref{ff_lor}) requires that $\alpha$ is constant along field 
lines, and hence field lines which leave the top or side of the box have 
$\alpha$ set to zero along their length. This approach deals with the missing boundary data on 
$\alpha$ and ensures that $\nabla \cdot \mathbf J =0$ is satisfied in the volume. The handling 
of missing boundary data is a source of uncertainty for the force-free 
modeling performed here, and in particular we expect the free energy to 
be decreased by the removal of currents implied by setting $\alpha$ to 
zero. In addition, images of the solution introduced by the periodicity
will affect the solution. The effect of this should be small when the 
computational domain is large so the region of interest is isolated from 
the boundaries. \\ 

\subsection{A Self-consistent Model of the Coronal Magnetic Field}
\label{intro_sc}

Force-free modeling encounters problems when applied to photospheric 
magnetogram data \citep{2009ApJ...696.1780D}. Firstly, the iterative 
methods used to solve Equations (\ref{ff_lor}) and (\ref{ff_ampere}) 
fail to strictly converge if large electric currents are present
in the boundary data \citep{2009ApJ...696.1780D}. Secondly, methods 
using the formally correct boundary conditions given in Section 
\ref{intro_forcefree} produce $P$ and $N$ solutions which differ 
both in terms of energy and field line connectivity. The second problem makes it difficult to estimate the 
energy of the region from the model because two solutions with different 
energy are obtained from the same magnetogram. These problems are in 
part due to noise in the data and uncertainties associated 
with the inversion of polarization measurements to give boundary magnetic field values, 
errors in the resolution of the 180-degree ambiguity in the direction of 
the field transverse to the line-of-sight (see Section \ref{modeling_data}), and 
errors introduced by the process of estimating photospheric values of 
$\alpha$. However the fundamental problem is likely the inconsistency 
between the force-free model and the magnetogram data \citep{1995ApJ...439..474M}.   

The self-consistency procedure developed by \citet{2009ApJ...700L..88W}
addresses the inconsistency between the boundary data and the force-free
model. The method iteratively modifies the values of $\alpha$ on the
boundary, which we denote $\alpha_0$, until the $P$ and $N$ solutions agree. 
The self-consistency procedure can also take into account the uncertainties 
on $\alpha_0$ which we denote $\sigma$. We refer to repeat applications of 
the method involving modifications of $\alpha_0$ as self-consistency `cycles'. 

Here we give a brief outline of the self-consistency procedure (see 
\citet{2009ApJ...700L..88W} or \citet{2011ApJ...728..112W} for a 
detailed explanation). A single self-consistency cycle can be broken 
into three steps. The first step involves constructing the $P$ and
$N$ solutions from the the boundary data $B_z(x,y)$ and $\alpha_0(x,y)$.
The Grad-Rubin method, an iterative method for solving Equations \eq{ff_lor}
and \eq{ff_ampere} \citep{gr} is used, with an implementation in code 
described by \citet{2007SoPh..245..251W}. In general the $P$ and $N$ solutions
obtained differ indicating the boundary data are inconsistent with
a force-free model. 

The second step of a self-consistency cycle involves constructing a 
second set of $\alpha$ values on the boundary, which we denote $\alpha_1$,
based on the connectivity of the $P$ and $N$ solutions. 
For a force-free field the parameter $\alpha$ is constant along field lines 
[an implication of Equation \eq{ff_lor}]. Closed magnetic field lines 
connect points in the boundary with $B_z>0$ to points with $B_z<0$, which 
defines a mapping between regions of $P$ polarity and $N$ polarity. The mapping defined
by the field lines of the $P$ solution maps values of 
$\alpha_0$ in the $P$ polarity to the $N$ polarity (since $\alpha$ is 
constant along field lines). This defines a second
set of $\alpha$ values over the $N$ polarity. Similarly the $N$ solution maps 
values of $\alpha_0$ from the $N$ polarity to the $P$ polarity, which defines
a second set of $\alpha$ values over the $P$ polarity. 
Combined the values of $\alpha_0$ mapped by the $P$  and $N$ solutions cover the entire lower 
boundary and we denote this second set of values $\alpha_1$. 
In general $\alpha_0(x,y) \ne \alpha_1(x,y)$ unless the boundary data are 
consistent with a force-free model. Uncertainties on $\alpha_1$, which we denote
$\sigma_1$, are constructed from $\sigma$ in the same way that $\alpha_1$ is
constructed from $\alpha_0$. 

The third step involves constructing a new set of $\alpha$ values on the lower boundary based on
$\alpha_0$ and $\alpha_1$. The self-consistency procedure treats $\alpha_0$ and 
$\alpha_1$ as two separate observations of $\alpha$ on the lower boundary 
and Bayesian probability is used to decide on the most probable value at 
each boundary point given the uncertainties $\sigma$ and $\sigma_1$ \citep{2003prth.book.....J}. 
Assuming the uncertainties $\sigma$ (and by construction also $\sigma_1$) 
are Gaussian uncertainties, the new distribution of $\alpha$ on the lower boundary 
is

\begin{equation}
\alpha_{\rm new} = \frac{\alpha_{0}/\sigma^2+\alpha_{1}/\sigma_{1}^2}{1/\sigma^2+1/\sigma_{1}^2 }.
\label{alpha_new}
\end{equation}

\noindent If all uncertainties are equal then $\alpha_{\rm new}$ is the average
of the two values at each point {\it i.e.} $\alpha_{\rm new} = (\alpha_0+\alpha_1)/2$. 
An important property of Equation \eq{alpha_new} is that $\alpha_{\rm new}$
is unchanged under a global rescaling of the uncertainties $\sigma$ and $\sigma_1$, {\it i.e.} 
Equation \eq{alpha_new} is unchanged by the replacement $\sigma \rightarrow C \sigma$,
where $C$ is a constant. Hence the absolute value of $\sigma$ at a point is not 
important: only the value with respect to other points is important. 

The process of constructing $P$ and $N$ solutions 
and then applying Equation \eq{alpha_new} (all three steps) is called a `self-consistency cycle'. 
This procedure is iterated until $\alpha_{0}=\alpha_{1}$ at all points in the 
boundary: a single force-free solution (a `self-consistent' solution)
is obtained. In \citet{2009ApJ...700L..88W} the self-consistency procedure 
was demonstrated in application to data from AR 10953, with uncertainties 
assumed equal at all boundary points, and in \citet{2011ApJ...728..112W} the calculation was 
repeated with uncertainties derived from the spectropolarimetric inversion. 
This paper follows the approach of \citet{2011ApJ...728..112W}.

%
%
\section{Data and Modeling}
\label{modeling}

\noindent Vector magnetogram data for active region AR 11029 are available 
from the ground based SOLIS Vector-SpectroMagnetograph (SOLIS/VSM), and from 
the Solar Optical Telescope SpectroPolarimeter on board the {\it Hinode} satellite ({\it Hinode}/SP). 
One SOLIS/VSM magnetogram per day is available for 24, 25, 26 and 28 October, 
and one {\it Hinode}/SP magnetogram per day is available from 24-29 October. All 
magnetograms are derived from observations of the Fe {\sc i} ($630{\rm nm}$)
multiplet which provides the vector magnetic field close to the level of 
the photosphere ($\approx 250\,{\rm km}$ above the base of the photosphere).

From the available data we use only the SOLIS/VSM magnetogram for 24 October
and the {\it Hinode}/SP magnetogram for 27 October to derive the relevant 
boundary conditions for force-free modeling, as outlined in Section \ref{intro_forcefree}. 
We exclude the data for 28 and 29 October because the active region was 
too close to the limb, where projections effects become prohibitive. We exclude the 
{\it Hinode}/SP data for 24-26 October because for these days most of the active region
lies outside the {\it Hinode}/SP field of view. We exclude the SOLIS/VSM data for 
25 and 26 October because the magnetograms appear to contain significant systematic 
errors and are not suitable for force-free modeling. In particular the component of the
magnetic field transverse to the line of sight shows a strongly preferred
direction which we attribute to the result of systematic error. 

The lower panel of Figure \ref{f1} shows the times of the acceptable magnetogram 
observations (solid vertical lines) compared with the flare history of the region. 
The magnetogram data requires $\approx 10$ minutes of integration time 
in the case of SOLIS/VSM and $\approx 60$ minutes of integration time in the 
case of {\it Hinode}/SP. The region first began flaring on 24 October producing 
several small {\it GOES} A-class and B-class, and the SOLIS/VSM magnetogram data are taken on this day. 
The flaring rate for the region peaked on 26 and 27 October and the {\it Hinode}/SP 
magnetogram data are obtained during this interval. \\

\subsection{SOLIS/VSM Data}
\label{modeling_data}

The National Solar Observatory SOLIS/VSM provides Milne-Eddington inverted vector magnetograms, 
and `Quicklook' data consisting of initial estimates of the 
vector magnetic field produced prior to Milne-Eddington inversion 
\citep{2006ASPC..358...92H}. Both full disk magnetograms and cropped data 
centered on individual active regions are available online\footnote{See \url{http://solis.nso.edu.} }. 
The online data provide the magnetic field vector close to the photosphere 
in terms of a line-of-sight component and a component transverse to the 
line of sight \citep{2006ASPC..358...92H}. The transverse component is 
further decomposed into a magnitude and direction defined by the azimuthal 
angle. The inversion is performed by the NSO and details are given 
in \citet{2006ASPC..358...92H}. 

The magnetogram for 24 October is obtained at $16 \colon 05\,{\rm (UT)}$ 
and is $236\times 236$ pixels in size with a spatial resolution of $1.1^{\prime \prime}$ per pixel. 
The magnetogram covers an area of $267^{\prime\prime}\times267^{\prime\prime}$
corresponding to $195\times196 \, {\rm Mm}$ on the
photosphere. The NSO also provides Quicklook data for the same day. Over regions of the 
photosphere where the magnetic field is weak there are gaps in the 
Milne-Eddington data where inversion in not performed due to a poor signal 
to noise ratio. We embed Quicklook data directly into the Milne-Eddington data to 
fill these gaps, following the advice provided by the NSO\footnote{See 
the introduction to SOLIS/VSM data at \url{http://solis.nso.edu/vsm/vsm_image_info3.html}.}. 

As explained in Section \ref{intro_forcefree} force-free modeling is 
performed in a locally flat Cartesian coordinate system defined on the 
photosphere, and we refer to this coordinate system as helioplanar coordinates. 
The magnetogram is provided on a uniform grid on a plane perpendicular to the line
of sight. We map the magnetogram to a helioplanar
system using the transformations given in \citet{1988SoPh..115..125V}. 
The helioplanar coordinate system is defined on a plane tangent to the solar surface at 
a point at the center of the active region. The axes of this system are 
oriented such that the $x$ and $y$ axes align with solar west and 
north respectively, and the $z$ axis is normal to the photosphere. Owing 
to projection effects after mapping, the grid on which the magnetic field 
is defined is no longer regularly spaced. We regrid the magnetogram onto 
a larger (in number of grid points) uniform grid with the same spatial resolution 
($1.1^{\prime \prime}$ per pixel) as the original data. Since the 
active region is close to disk center on 24 October the projection effects 
are small.

The spectropolarimetrically derived azimuthal angle contains 
an inherent $180^{\circ}$ ambiguity \citep{2004ASSL..307.....L} which
must be resolved to determine the magnetic field components ($B_x$, $B_y$ and $B_z$)
needed to derive boundary conditions for force-free modeling.
Several procedures exist for resolving the $180^{\circ}$ ambiguity 
[see \citet{2006SoPh..237..267M} and references therein], and 
here we use the `minimum energy' method developed by 
\citet{1994SoPh..155..235M}. This method determines the azimuthal angle by requiring 
the ambiguity-resolved field to simultaneously minimize 
$|\nabla \cdot \mathbf B|$ and $|\mathbf J|^2$ over the magnetogram. 
The optimal magnetic field configuration is found using the general 
optimization procedure of simulated annealing \citep{1992nrfa.book.....P}. 
We note that the NSO/SOLIS provides ambiguity-resolved data using a different 
resolution method \citep{2005ApJ...629L..69G}, but we have chosen to perform this step 
again using the minimum energy procedure.

The boundary conditions for force-free modeling are 
the component of the magnetic field normal to the photosphere,
which corresponds to the $B_z$ values in the regridded magnetogram, and    
the force-free parameter $\alpha$ on the photosphere derived from the values of 
$B_x$ and $B_y$ in the magnetogram using Equation (\ref{alpha0}). 
The partial derivatives in Equation (\ref{alpha0}) are approximated using 
centered differencing \citep{1992nrfa.book.....P}. The boundary values on 
$\alpha$ are set to zero at points where $|B_z| < 0.01 \times \mbox{max}(|B_z|)$ 
because the ratio $J_z/B_z$ is inaccurate in regions where $B_z$ is small. 

\begin{figure}
\centerline{\includegraphics[scale=0.85]{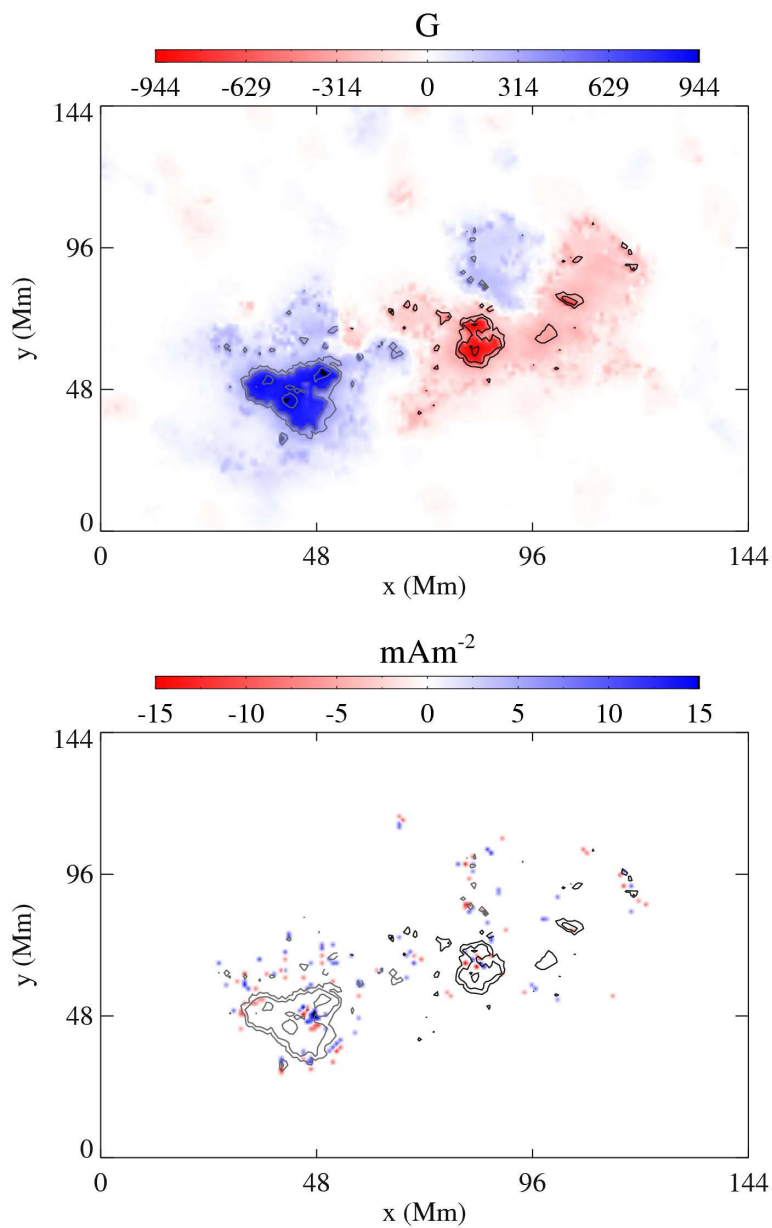}}
\caption{The central part of part of the SOLIS/VSM vector magnetogram data
for 24 October. The upper panel shows $B_z$  
and the lower panel shows $J_z$. The data are colored blue when 
positive, white when zero, and red when negative. Contours of $B_z$ are 
superimposed on the data, with contours separated by intervals of $300\,{\rm G}$. 
The contours are black where $B_z<0$ and light gray where $B_z>0$. 
The values of $B_z$ (upper panel) have been saturated at
$\pm 944 \, {\rm G}$ with a maximum value of $961 \, {\rm G}$.
The values of $J_z$ (lower panel) have been saturated at
$\pm 15 \, {\rm mAm^{-2}}$ with a maximum value of $23 \, {\rm mAm^{-2}}$. 
In the lower panel only points where $|\alpha_0|/\sigma$ is three 
standard deviations above the mean are shown.}
\label{f2}
\end{figure}

\begin{figure}
\centerline{\includegraphics[scale=0.85]{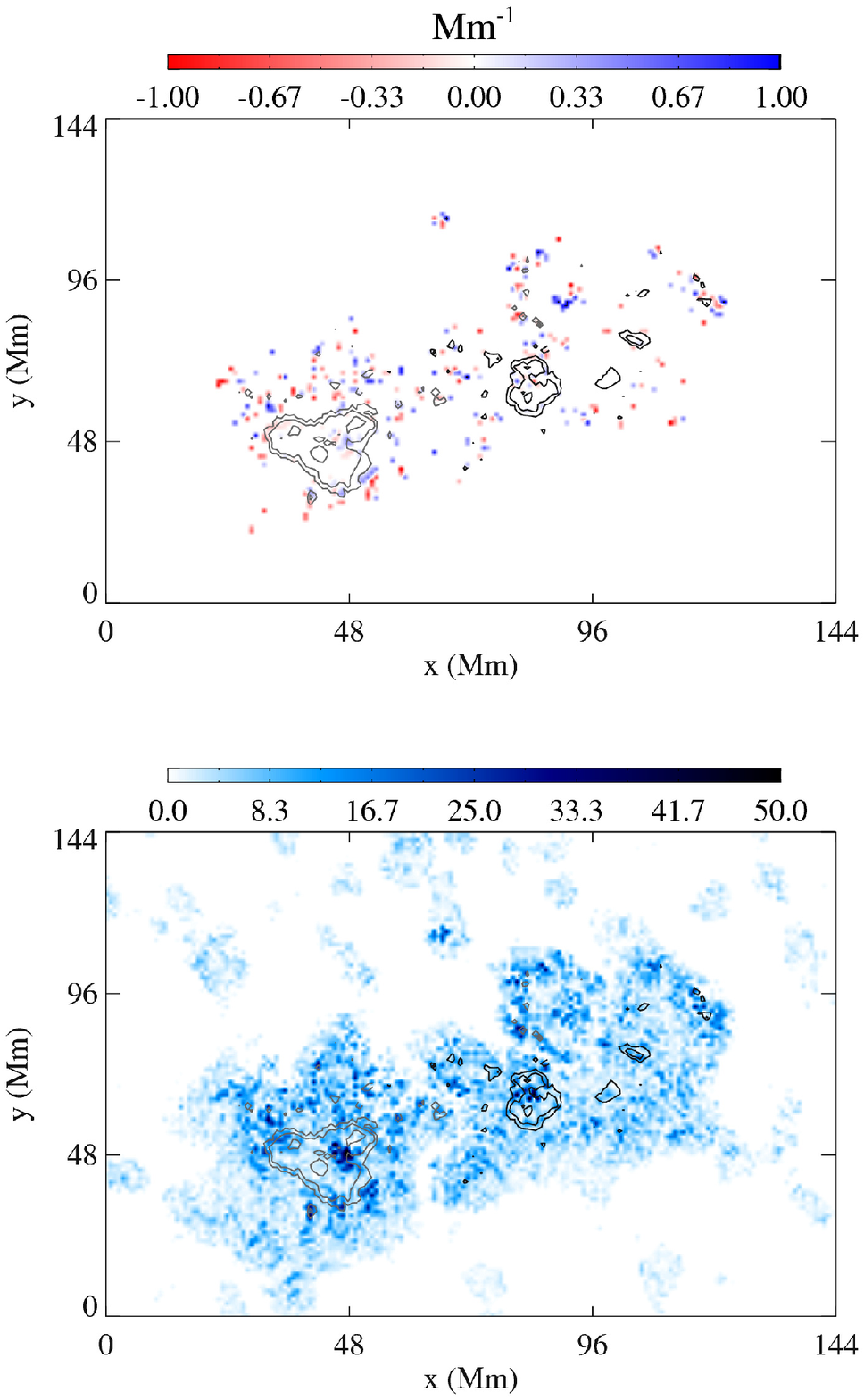}}
\caption{Boundary conditions on the force-free parameter obtained from
the SOLIS/VSM vector magnetogram data for 24 October. 
The upper panel shows $\alpha_0$ and the lower panel shows the signal to 
noise ratio $|\alpha_0|/\sigma$. The data are colored blue when positive, white when zero, and red 
when negative. Contours of $B_z$ are superimposed on the data, with 
the contours separated by $300\,{\rm G}$. 
The contours are black where $B_z<0$ and light gray where $B_z>0$. 
The values of $\alpha_0$ (upper panel) have been saturated at
$\pm 1\,{\rm Mm^{-1}}$ with a maximum value of $3.7\,{\rm Mm^{-1}}$.
The values of $|\alpha_0|/\sigma$ (lower panel) have been saturated at
$50$ with a maximum value of $67$. In the upper panel only points where 
$|\alpha_0|/\sigma$ is three standard deviations above the mean are shown.}
\label{f3}
\end{figure}

The final boundary conditions after projection and regridding are shown 
in Figures \ref{f2} and \ref{f3}. Figure \ref{f2} shows the boundary 
conditions for $B_z$ (upper panel), and the boundary conditions for $J_z$ 
(lower panel). Figure \ref{f3} shows the force-free parameter $\alpha_0$,
and the signal to noise ratio $|\alpha_0|/\sigma$ for $\alpha_0$ 
where $\sigma$ represents the approximate uncertainty in the $\alpha_0$ 
value estimated using the $B_z$ boundary value, as explained in 
Section \ref{uncert}. In both Figures \ref{f2} and \ref{f3}, 
the data saturate at a particular threshold value which is below the maximum in the data, 
by which we mean all points above the threshold appear with the same color. 
This is done to provide a good contrast between large and small data 
values, and to diminish the appearance of outliers. For the panels showing $J_z$ and 
$\alpha_0$ only points where the signal to noise ratio $|\alpha_0|/\sigma$ is three 
standard deviations above the mean signal to noise ratio are represented, 
points where $|\alpha_0|/\sigma=0$ are excluded when computing the mean and standard
deviation. This criteria is chosen because it is independent of the magnitude
of $\sigma$ which is arbitrary for the SOLIS/VSM data (see Section \ref{uncert}).
Only the central portion of the data is shown in the two panels because
the boundary data values ($B_z$, $J_z$ and $\alpha_0$) outside the central region are 
comparatively small and are not visible with the chosen scale. 

\subsection{{\it Hinode}/SP Magnetogram Data}
\label{modeling_data2}

The {\it Hinode}/SP vector magnetogram data for 27 October is derived
from spectropolarimetric data from the Solar Optical
Telescope/SpectroPolarimeter aboard the {\it Hinode} satellite
\citep{2007SoPh..243....3K, 2008SoPh..249..167T}.  The `fast-scan' map obtained at
$10\colon45\,{\rm (UT)}$ is spatially sampled at $0.32^{\prime \prime}$, and
covers a relatively small $360^{\prime \prime} \times 164 ^{\prime
\prime}$ field of view. The data are inverted using the
High Altitude Observatory (HAO) Milne-Eddington inversion code (\opencite{1987ApJ...322..473S};
\opencite{1990ApJ...348..747L}; \opencite{1993ApJ...418..928L}), which has 
been modified for {\it Hinode}/SP data (revised code courtesy B. Lites, HAO/NCAR).
Ambiguity resolution is performed using an implementation\footnote{Available at \url{http://www.cora.nwra.com/AMBIG/}.} 
of the minimum energy procedure \citep{2009ASPC..415..365L}. 

For the {\it Hinode}/SP data for 27 October uncertainties in the three 
components of the magnetic field $(\sigma_x,\sigma_y,\sigma_z)$ are 
derived during the Milne-Eddington inversion. These are based on the 
uncertainty in the $\chi^2$ fit of the synthetic Stokes 
profiles to the observed Stokes profiles. The uncertainties are then 
propagated through the projection, regridding and ambiguity resolution 
to derive uncertainties for $\alpha_0$. These uncertainties are a measure 
of how well the artificial Milne-Eddington profiles fit the observed 
profiles. They do not take into account the approximation of the 
Milne-Eddington atmosphere or systematic errors in the observations of the Stokes profiles. 
Therefore they may be considered lower bounds to the real uncertainties.
Some parameter-dependence is introduced into the method 
of simulated annealing by the choice of a `cooling schedule' \citep{1992nrfa.book.....P}.
We note that some data points are particularly sensitive to the choice of schedule,
but overall the large scale structure of the ambiguity-resolved magnetogram
appears independent of the choice. Large uncertainties are assigned 
to points which depend strongly on the choice of cooling schedule
to reflect this parameter dependence. 

The resulting vector magnetic components and uncertainties from the
{\it Hinode}/SP data are re-mapped to a helioplanar $1^{\prime
\prime}$ grid. The {\it Hinode}/SP field of view is small and only covers
the main portion of the active region. As explained in Section \ref{intro_forcefree} 
when performing force-free modeling it is important to isolate the 
region of interest from the boundaries to limit the influence of the boundaries
on the solution. As with other studies involving {\it Hinode}/SP data
({\it e.g.} \citealt{2011ApJ...728..112W}; \citealt{2009ApJ...696.1780D})
the field of view is expanded by adding line-of-sight data from
the Michelson Doppler Imager on the {\it Solar and Heliospheric
Observatory} \citep{1995SoPh..162..129S} around the {\it Hinode}/SP data. 
A map of $B_{z}$ is constructed using a potential field matching the observed 
line-of-sight magnetic field values in the boundary, and this is also remapped 
onto a $1^{\prime \prime}$ grid. After coalignment between the two data sources,
an apodizing function is applied at the boundaries of the two data sets to ensure a smooth 
transition between data values. This procedure follows \citet{2011ApJ...728..112W}, 
and improves upon a method used in \citet{2009ApJ...696.1780D}.  
The final size of the boundary data is $440\times300$ pixels, covering a 
region on the photosphere of size $320 \times 220 \,{\rm Mm}$.

\begin{figure}
\centerline{\includegraphics[scale=0.85]{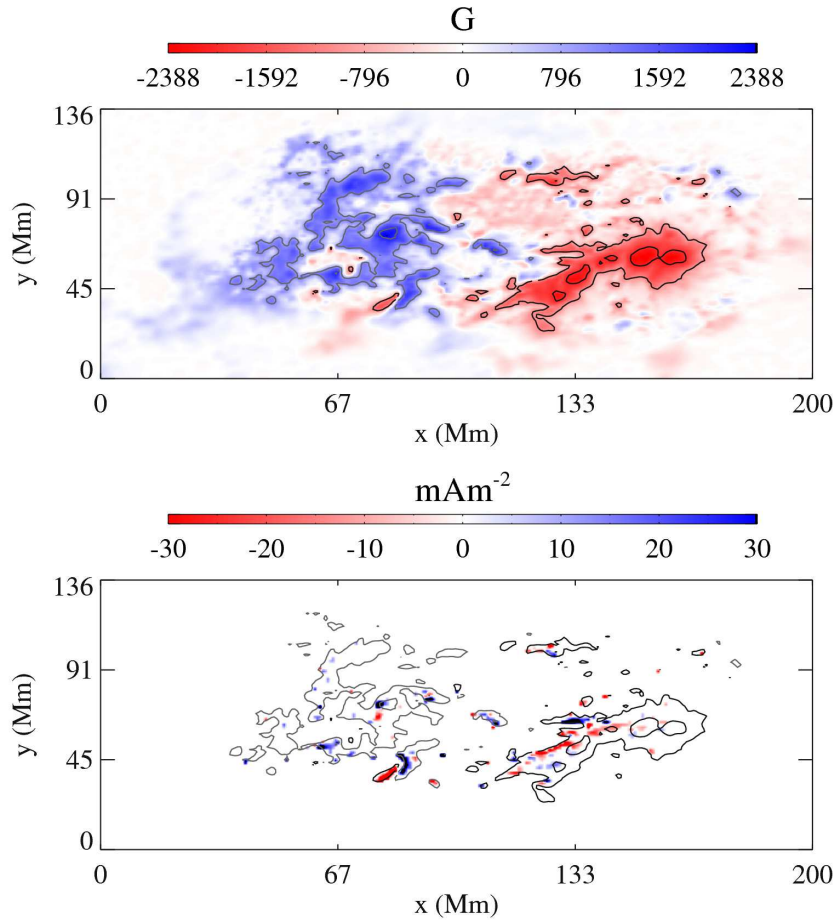}}
\caption{The central part of the {\it Hinode}/SP vector magnetogram data
for 27 October. The upper panel shows $B_z$  
and the lower panel shows $J_z$. The data are colored blue when 
positive, white when zero, and red when negative. Contours of $B_z$ are 
superimposed on the data, with contours separated by intervals of $1000\,{\rm G}$. 
The contours are black where $B_z<0$ and light gray where $B_z>0$. 
The values of $B_z$ (upper panel) have been saturated at
$\pm 2448 \, {\rm G}$ with a maximum value of $2388 \, {\rm G}$.
The values of $J_z$ (lower panel) have been saturated at
$\pm 30 \, {\rm mAm^{-2}}$ with a maximum value of $144 \, {\rm mAm^{-2}}$. 
In the lower panel only points where $|\alpha_0|/\sigma$ is three 
standard deviations above the mean are shown.} 
\label{f4}
\end{figure}

\begin{figure}
\centerline{\includegraphics[scale=0.85]{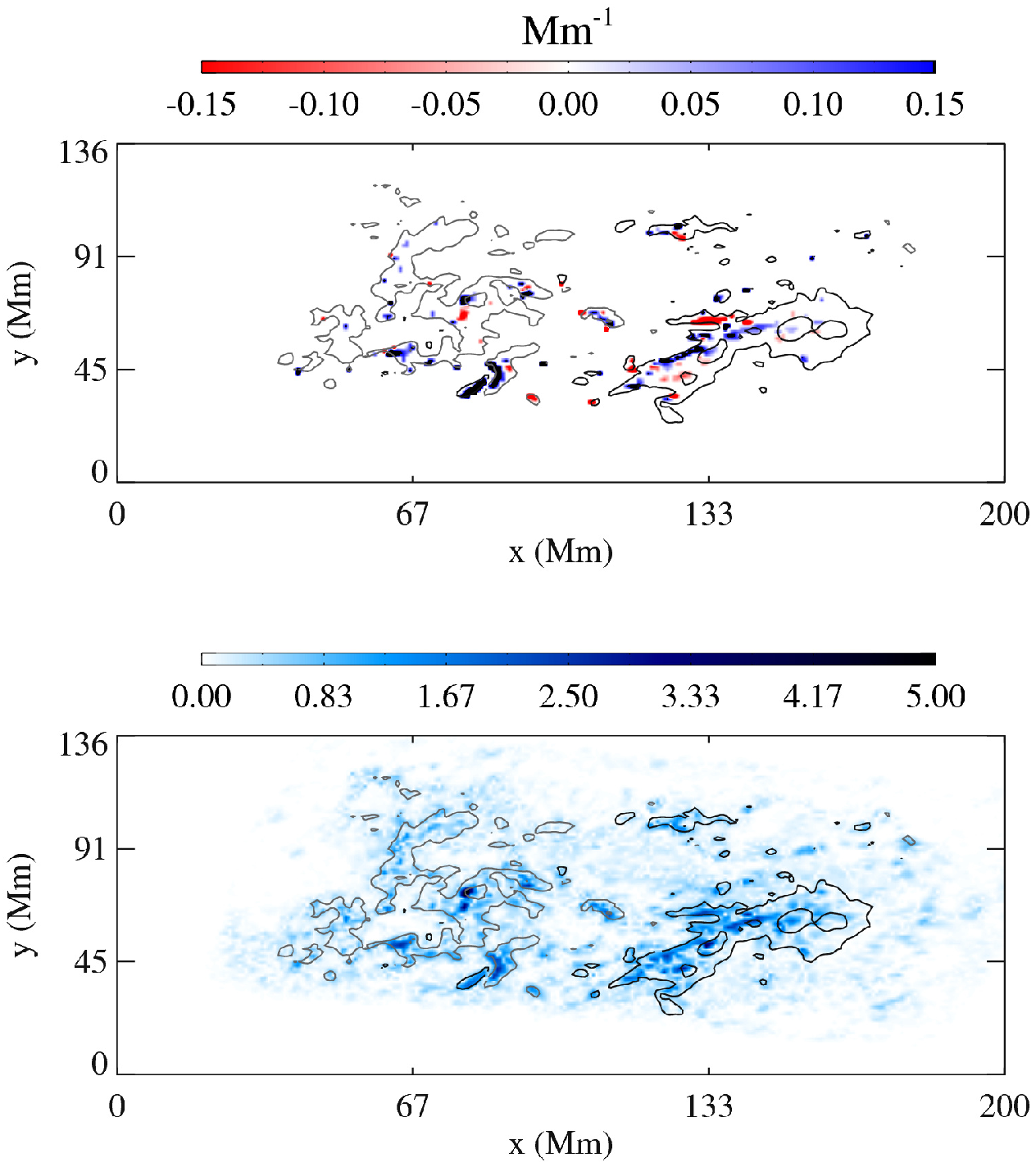}}
\caption{Boundary conditions on the force-free parameter obtained from
the {\it Hinode}/SP vector magnetogram data for 27 October. 
The upper panel shows $\alpha_0$ and the lower panel shows the signal to 
noise ratio $|\alpha_0|/\sigma$. The data are colored blue when positive, white when zero, and red 
when negative. Contours of $B_z$ are superimposed on the data, with 
the contours separated by $1000\,{\rm G}$. 
The contours are black where $B_z<0$ and light gray where $B_z>0$. 
The values of $\alpha_0$ (upper panel) have been saturated at
$\pm 0.15\,{\rm Mm^{-1}}$ with a maximum value of $1.9\,{\rm Mm^{-1}}$.
The values of $|\alpha_0|/\sigma$ (lower panel) have been saturated at
five with a maximum value of $5.4$. In the upper panel only points where 
$|\alpha_0|/\sigma$ is three standard deviations above the mean are shown.}
\label{f5}
\end{figure}

The data are shown in Figures \ref{f4} and \ref{f5}. The vertical magnetic
field $B_z$ (upper panel) and vertical current density $J_z$ (lower panel)
are shown in Figure \ref{f4}. The presentation of the data follows
the format used in Figures \ref{f2} and \ref{f3}. 
Approximate uncertainties $\sigma$ in the $\alpha_0$ values are provided by the 
spectropolarimetric inversion, for the {\it Hinode}/SP data points. 
Figure \ref{f5} shows the force-free parameter $\alpha_0$ (upper panel) 
and the signal-to-noise ratio $|\alpha_0|/\sigma$ (lower panel). 
For the panels showing $J_z$ and $\alpha_0$ only points where the signal 
to noise ratio $|\alpha_0|/\sigma$ is three standard deviations above 
the mean signal to noise ratio are represented, points 
where $|\alpha_0|/\sigma=0$ are excluded when computing the mean and standard
deviation. Only the central portion of the data is shown in the two panels because
the boundary data values ($B_z$, $J_z$ and $\alpha_0$) outside the central region are 
comparatively small and are not visible with the chosen scale. 

\subsection{Flux Balance}
\label{f_balance}

The integral of $B_z$ over the magnetogram defines the net magnetic flux across 
the observed region of the photosphere. If the net flux is zero then the magnetogram is said to
be flux balanced.  Flux balance is not essential for the force-free 
modeling procedure used here, but if the flux is unbalanced a greater 
number of field lines leave the top and sides of the computational 
domain in order to satisfy $\nabla \cdot \mathbf B=0$ globally. As 
discussed in Section \ref{intro_forcefree} boundary conditions are not 
known on the top and side boundaries of the domain and the force-free 
code sets $\alpha=0$ along field lines which leave the domain via these 
boundaries \citep{2007SoPh..245..251W}. The currents and therefore the 
free energy  tend to be reduced by this procedure. The effect of 
the boundaries on the energy estimates is likely to be less pronounced if the 
magnetogram is flux balanced.

A practical measure of flux balance is the relative flux balance, defined as

\begin{equation}
\Phi_{B} = \frac{\int B_z dA}{\int |B_z| dA}
\label{rel_flux}
\end{equation} 

\noindent where the integrals are over the magnetogram area. The integrals in 
Equation (\ref{rel_flux}) are evaluated numerically using the two-dimensional trapezoidal 
rule. The results are listed in the right-hand column in Table \ref{t1}. 

We consider a $\Phi_B$ of less than $1\%$ acceptable. The {\it Hinode}/SP 
magnetogram for 27 October meets this requirement, but the SOLIS/VSM magnetogram 
for 24 October does not. To flux balance the SOLIS/VSM data a buffer 
region of constant $B_z$ is added around the border of the magnetogram. The magnitude 
of the field in the buffer region satisfies $|B_z| < 0.01 \times \mbox{max}(|B_z|)$, 
and hence the buffer region is considered weak field by our previous definition 
(see Section \ref{modeling_data}). In the buffer the magnitude of $B_z$ is 
$9.3\, {\rm gauss \, (G)}$ (the weak field threshold is $9.6 \,{\rm G}$), and 
the value of $\alpha_0$ is set to zero everywhere in the buffer. The total 
size of the magnetogram with the buffer added is $270\times270$ grid points, and the resolution 
of the data remains the same ($1.1^{\prime\prime}$ per pixel).
The final size of the region including the flux buffer is $222\times222 \,{\rm Mm}$. 

\subsection{Phenomenological Uncertainties for SOLIS/VSM Data}
\label{uncert}

Uncertainties are provided with the {\it Hinode}/SP data based on 
spectropolarimetric inversion of the data (as explained in Section \ref{modeling_data2}), but 
are not available for the SOLIS/VSM data for 24 October. 
Instead we make a simple estimate for the uncertainties based on the 
magnitude of $B_z$ at each boundary point. Specifically we introduce
a phenomenological uncertainty defined by 

\begin{equation}
L_x \sigma(x,y) = \left \{\begin{array}{rl} 
0.01 & \mbox{ for the flux buffer region;} \\ 
B_{\rm m}/|B_z(x,y)| & \mbox{where $|B_z/B_{\rm m}| > 0.01$} \\
100 & \mbox{where $|B_z/B_{\rm m}| < 0.01$} \\
\end{array} \right. 
\label{errors_solis}
\end{equation}

\noindent Equation (\ref{errors_solis}) is written in non-dimensional
units. The magnetic field is scaled by $B_{\rm m}=\mbox{max}(|B_z|)$, the maximum value of 
$|B_z|$, and $\sigma$ is scaled by $L_x$, the transverse length of
the magnetogram. Points in the flux buffer are assigned a small constant uncertainty, 
which ensures $\alpha_0=0$ for this region during the self-consistency procedure. 
A relatively large uncertainty ($L_x \sigma=100$, in non-dimensional units) is assigned 
to points in weak field regions where $\alpha_0$ has been set to zero 
(see Section \ref{modeling_data}). 

The phenomenological scaling $\sigma(x,y) \sim 1/|B_z(x,y)|$ for the 
SOLIS/VSM data region is based on inspection of the inversion-based 
uncertainties for the {\it Hinode}/SP data for 27 October.

\begin{figure}
\centerline{\includegraphics[width=12cm,height=8cm]{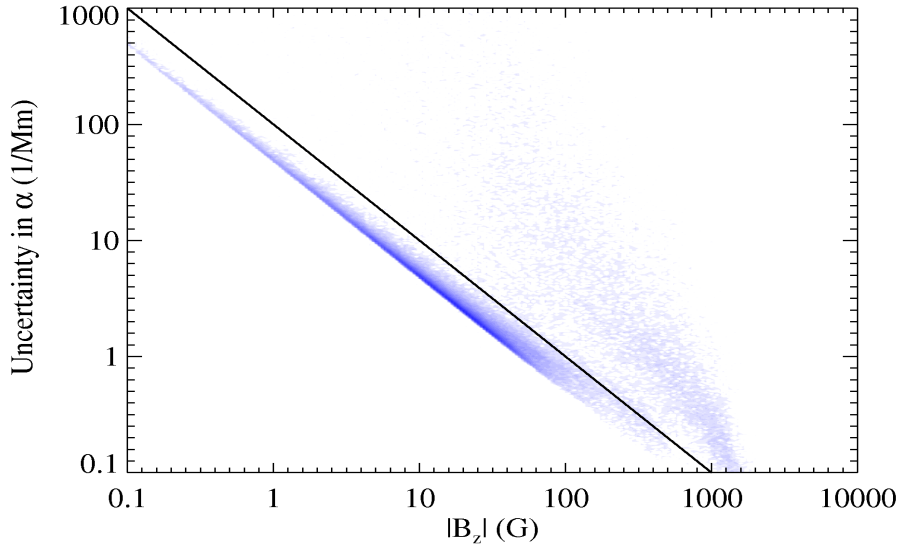}}
\caption{The relationship between the uncertainties on the boundary values 
of the force-free parameter derived from the spectropolarimetric inversion  
and $|B_z|$ for the {\it Hinode}/SP data for 27 October. The solid line is 
to illustrate $\sigma \sim 1/|B_z|$ dependence and the position of the
line has been chosen arbitrarily. There is a noticeable departure from
the $\sigma \sim 1/|B_z|$ scaling for large values of $|B_z|$.}
\label{f6}
\end{figure}

Figure \ref{f6} shows $P(\sigma,|B_z|)^{1/4}$, 
where $P(\sigma,|B_z|)$ is the distribution of 
the inversion-based uncertainties for the {\it Hinode}/SP
data for 27 October over field strength and uncertainty in $\alpha_0$ 
{\it i.e.} the fraction of points with field value $|B_z|$ and uncertainty 
$\sigma$. We have applied a nonlinear scaling to the distribution so 
that strong field regions with small values of $P(\sigma,|B_z|)$ are visible.
The solid line shows $\sigma = 100\,{\rm Mm^{-1}}/|B_z|$ with $B_z$ in
gauss, to illustrate $\sigma \sim 1/|B_z|$ dependence for comparison. 
The choice of the factor $100 \, {\rm Mm^{-1}}$ is arbitrary.      
The figure shows that the simple phenomenological scaling implied by 
Equation \eq{errors_solis} is accurate for a majority of data points, although
their is notable departure for large values of $|B_z|$. 

It should be noted that, according to Equation \eq{alpha_new}, the exact 
magnitude of the uncertainties is not important for the self-consistency
averaging procedure [see discussion following Equation \eq{alpha_new}]. 
Only the relative size of uncertainties at different points is important. 

Although, due to the departure from the $\sigma \sim 1/|B_z|$ scaling for 
large values of $|B_z|$ we expect that the uncertainties in strong field
regions are underestimated relative to weak field regions.  

\subsection{Force-free Modeling Using SOLIS/VSM and {\it Hinode}/SP Data}
\label{modeling_forcefree}

\noindent Force-free solutions are calculated using the code of
\citet{2007SoPh..245..251W} and the self-consistency procedure of 
\citet{2009ApJ...700L..88W}. The boundary 
conditions derived from the SOLIS/VSM and {\it Hinode}/SP vector magnetogram data
described in Sections \ref{modeling_data} and \ref{modeling_data2} are used. 
The initial magnetic field for the Grad-Rubin iteration is a potential 
field calculated using a Fourier solution to Poisson's equation. 
The computational domain used is a uniform three dimensional Cartesian grid with transverse dimensions ({\it i.e.}\ the 
$x$ and $y$ dimensions) matching the dimensions of the magnetogram which 
forms the lower boundary at $z=0$. The vertical dimension ({\it i.e.} $z$) is 
$128$ points high for the SOLIS/VSM data and $200$ points high for the {\it Hinode}/SP 
data. The resulting dimensions of the grids are $270 \times 270 \times 128 $ 
points for 24 October, and $440 \times 300 \times 200$ points for 27 October. 
The grid spacing is uniform in each direction. We summarize the size, maximum 
magnetic field, and flux balance of the data in Table \ref{t1}. 

\begin{table}[!h]
\caption{Summary of magnetogram data/computational data domain sizes, 
field strength parameters, and relative flux balance $\Phi_B$ 
(see Equation (\ref{rel_flux})) for the computational models of AR 11029.}
\begin{tabular}{l  l  l  l  l  l}
\hline
\hline             
 Data source & Date &  $N_x \times N_y \times N_z $ & $L_x \times L_y \times L_z $& $B_{\rm m}=\mbox{max}(|B_z|)$ & $\Phi_{B}$ \\  
             &      &                               & ($10^{8}\, {\rm m})$          & (G)          & ($\%$) \\           
\hline
SOLIS/VSM & Oct. 24 & $236 \times 236 \times 128$ & $2.0 \times 2.0 \times 1.1 $ &  962 & 11 \\   
SOLIS/VSM (buffered)   & Oct. 24   & $270 \times 270 \times 128$ & $2.2 \times 2.2 \times 1.1$ & 962 & 0.02 \\
{\it Hinode}/SP        & Oct. 27   & $440 \times 300 \times 200$ & $3.2 \times 2.2 \times 1.5$ & 2448 & 0.4 \\
\end{tabular}
\label{t1}
\end{table}

%
%
\section{Results}
\label{results}

\noindent This section presents the results of the force-free modeling
of active region AR 11029. Section \ref{res_24} discusses the results
when the force-free code is applied to the SOLIS/VSM data for 24 October to construct
$P$ and $N$ solutions without using the self-consistency procedure. 
Section \ref{sc_24} discusses the application of the self-consistency procedure 
to the same data. Section \ref{res_27} discusses the application of the
self-consistency procedure to the {\it Hinode}/SP data for 27 October. 

\subsection{Construction of $P$ and $N$ Solutions Directly from Magnetogram 
Data for 24 October}
\label{res_24}

\noindent We construct $P$ and $N$ solutions for AR 11029 directly from the
magnetogram data for 24 October. In this section we do not 
apply the self-consistency procedure. To construct the $P$ and $N$ solutions
50 Grad-Rubin iterations are used beginning with a potential field constructed
from $B_z$.

\begin{figure}[!h]
\centerline{\includegraphics[width=12cm,height=8cm]{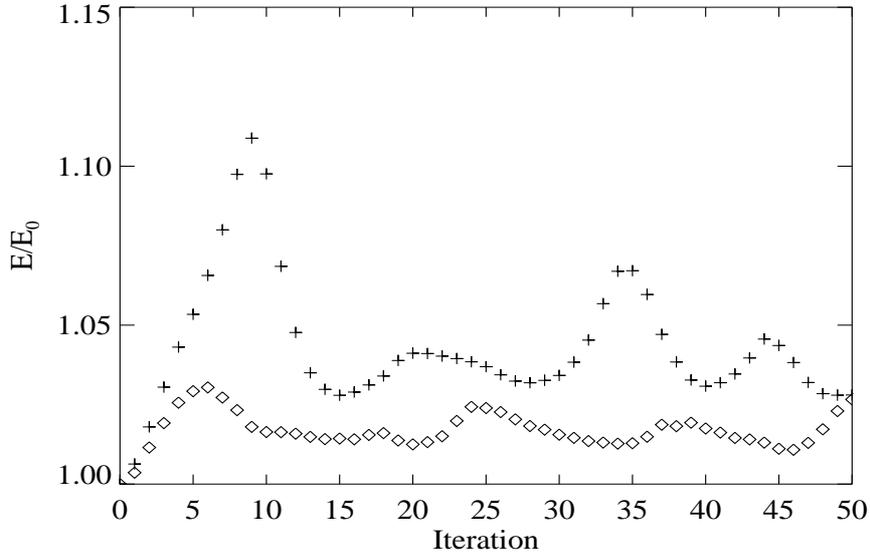}}
\caption{Results for the construction of $P$ and $N$ solutions directly from
the SOLIS/VSM magnetogram data for 24 October. The total magnetic energy
is shown (as a fraction of the energy of the potential field) for the $P$ 
solution (plus signs) and the $N$ solution (diamonds) over $50$ Grad-Rubin 
iterations starting from a potential field. The oscillations in the energy 
occur because of large localized currents in the boundary data.}
\label{f7}
\end{figure}
     
In application to the magnetogram data the Grad-Rubin procedure does not
strictly converge for the $P$ and $N$ solutions. The energy of the 
magnetic field at each iteration provides a way to examine 
the convergence. Figure \ref{f7} shows the magnetic 
energy, $E$, in units of the energy of the potential field, $E_0$, over 
successive Grad-Rubin iterations for both the $P$ and $N$ solutions. 
The energy of the two solutions behaves similarly. Initially the total 
energy increases beyond the energy of the potential field and 
then begins oscillating after about $15$ iterations for the $P$ 
solution and after about ten iterations for the $N$ solution. This 
oscillatory behavior is typical in situations where the Grad-Rubin 
method does not strictly converge, and may be attributed to large currents at 
certain points in the boundary. This is confirmed  
by inspecting visualizations of field lines at successive 
Grad-Rubin iterations. For the $P$ and $N$ solutions for the 24 October 
data we observe that the longest field lines which stretch high into
the computational volume (and carry significant currents) change 
over successive iterations. The field in the lower part of the volume is static, 
and the field in the middle of the volume displays small oscillations.   

We estimate the energy of the $P$ and $N$ solutions by averaging the
energy over several Grad-Rubin iterations. 
After 15 Grad-Rubin iterations the $P$ solutions stops changing systematically 
and begins oscillating so we estimate the energy of the $P$ solution by taking the average energy
of the field over successive Grad-Rubin iterations starting from iteration 15
[{\it i.e.} we take $E_{P}= {\rm mean}(E_{15},E_{16},...,E_{50})$, where 
$E_{i}$ is the energy of the $P$ solution after $i$ Grad-Rubin iterations].    
We do the same for the $N$ solution, except we start the average from iteration
ten [{\it i.e.} $E_{N}={\rm mean}(E_{10},...,E_{50})$]. We also compute the free 
energy $E_{\rm f}=E-E_0$, for $E=E_P$ and $E=E_N$. The results are listed in Table \ref{t2}. 

\begin{table}
\caption{Energy estimates for the $P$ and $N$ solutions constructed directly 
from the SOLIS/VSM magnetogram data for AR 11029 on 24 October.}
\begin{tabular}{l  l  l  l  l }
\hline
\hline
 Solution & $E$                      & $E_0$                   & $E_{\rm f}=E-E_0$  \\  
   & ($10^{31}\,{\rm erg}$ ) & ($10^{31}\,{\rm erg}$) & ($10^{30}\,{\rm erg}$) \\
\hline
$P$  & $9.98$ & $9.62$ &  $3.6 $   \\
$N$  & $9.77$ & $9.62$ &  $1.5 $  \\

\end{tabular}
\label{t2}
\end{table}

\begin{figure}[!h]
\centerline{\includegraphics[scale=0.85]{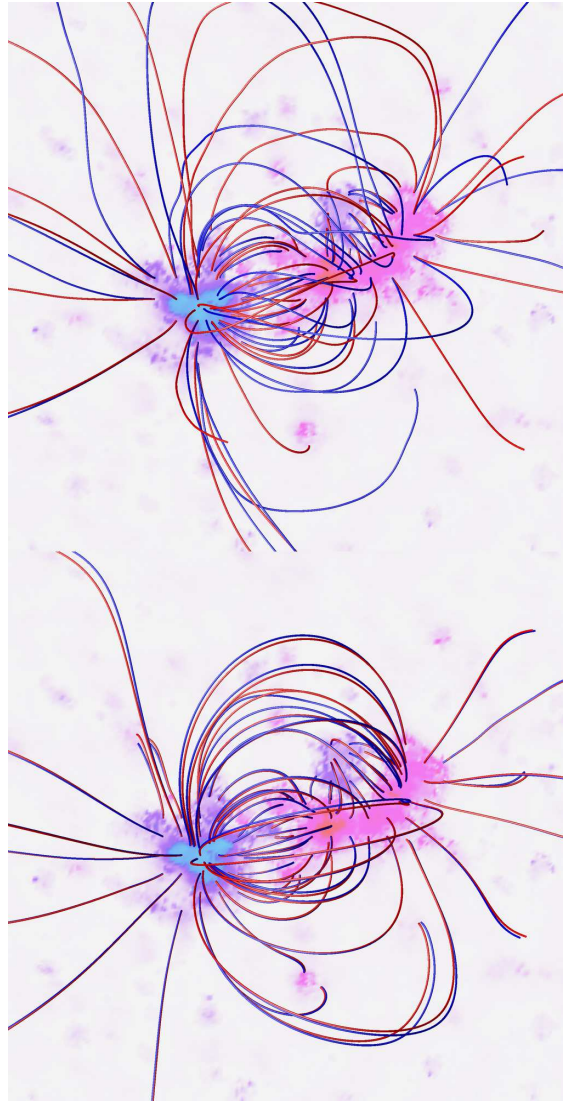}}
\caption{Nonlinear force-free models of the coronal magnetic field 
constructed from the SOLIS/VSM vector magnetogram data for 24 October. 
The $P$ solution (blue field lines) and $N$ solution (red field lines) are shown 
viewed looking down on the photosphere. The upper panel shows the 
two solutions constructed from the original vector magnetogram boundary data (with
a flux buffer added) using 20 Grad-Rubin iterations. The lower panel shows the 
solutions after ten self-consistency cycles with $20$ Grad-Rubin iterations per cycle. 
The photosphere is colored blue where $B_z>0$ and red where $B_z<0$. 
This figure illustrates the achievement of self-consistency: the $P$ and $N$ solutions 
in the upper panel are dissimilar but in the lower panel are very similar.}
\label{f8}
\end{figure}

The upper panel of Figure \ref{f8} shows a visualization of the magnetic field after 
20 Grad-Rubin iterations (the lower panel is discussed in Section \ref{sc_24}).
The magnetic field lines for the $P$ solution (blue curves) and the $N$ 
solution (red curves) are displayed in the upper panel. The fields are 
viewed looking down on the photosphere, which is colored blue where $B_z>0$ 
and red where $B_z<0$. The panel shows some qualitative difference between 
the $P$ and $N$ solutions.

The visualization of the field lines also reveals the extent to which the 
missing information on the boundaries influences the solution. It can be 
seen from the upper panel of Figure \ref{f8} that both solutions have field 
lines which cross the side boundaries. Many field lines connected to the trailing positive 
spot leave the box. However, a significant part of the active region field 
involves closed field lines. It is unlikely the boundary has a large
influence on the final energy. 

The results in this section illustrate difficulties associated with 
obtaining estimates for the total and free energy using the nonlinear 
force-free model. The Grad-Rubin iteration does not strictly converge, due
to strong currents, which introduces a degree of arbitrariness in the 
energy estimate since the energy depends on the choice of when to stop 
the iteration process. Compared with the total energy the 
oscillations in the total energy are small, but these oscillations 
in the total energy translate into large oscillations 
in the free energy, because $E \gg E_{\rm f}$. The force-free energy estimates 
may be regarded as approximate (order of magnitude) estimates of the 
free energy. Based on the results in Table \ref{t2} we estimate that the free energy is 
$\approx 3 \times 10^{30} \, {\rm erg}$. 

\subsection{Construction of Self-consistent Solutions for 24 October}
\label{sc_24}

The self-consistency procedure is applied to the SOLIS/VSM vector 
magnetogram boundary data for 24 October. Initially, we 
use ten self-consistency cycles with $20$ Grad-Rubin iterations 
per cycle ($P$ and $N$ solutions are constructed using $20$ Grad-Rubin 
iterations, and this is repeated for ten self-consistency cycles, with the $\alpha_0$-values 
updated using Equation \eq{alpha_new} at the end of each cycle). 

\begin{figure}
\centerline{\includegraphics[width=12cm,height=8cm]{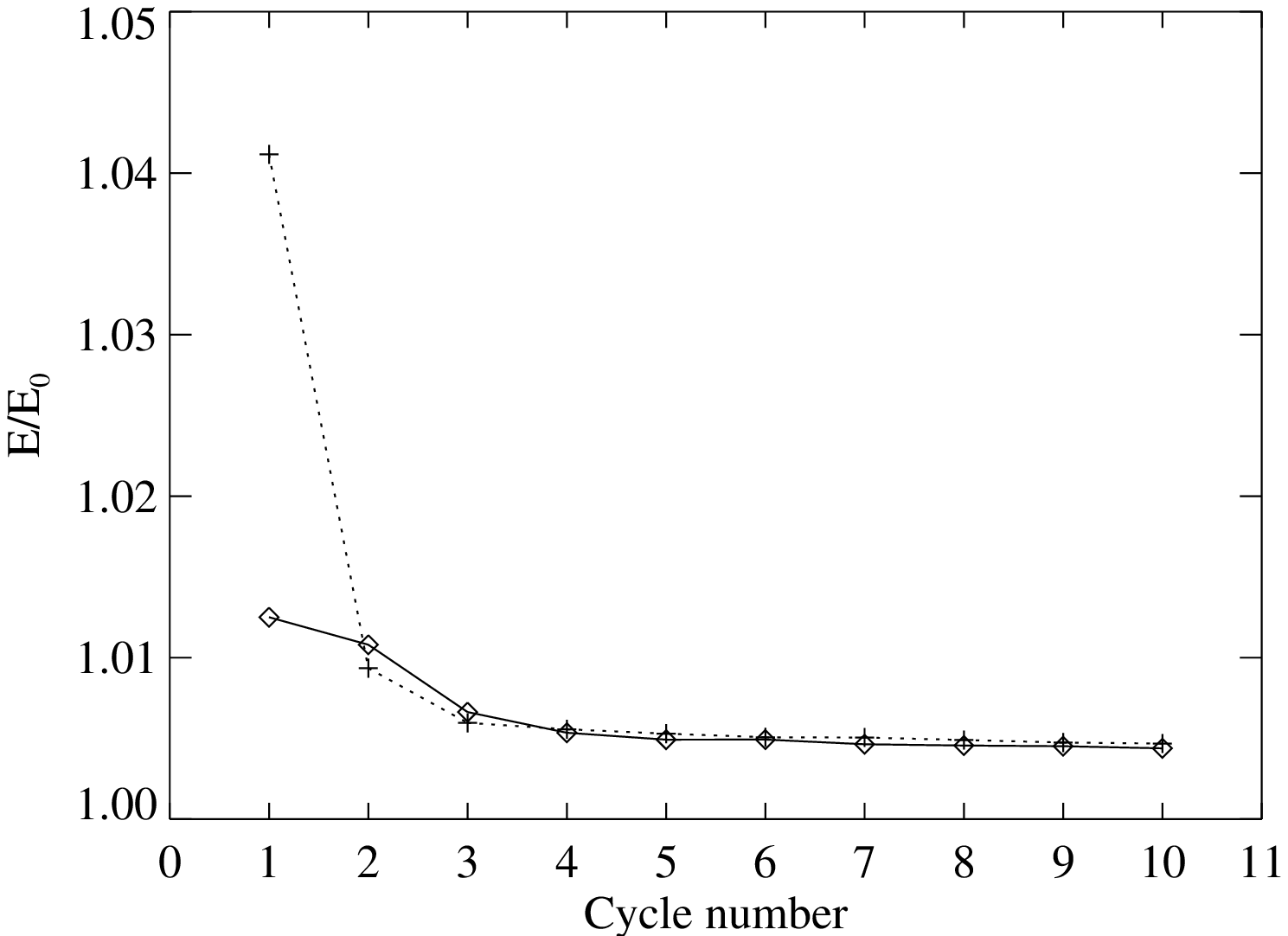}}
\caption{Results for the application of the self-consistency procedure to the
SOLIS/VSM vector magnetogram data for 24 October. The energies of 
the $P$ and $N$ solutions (as a fraction of the energy of the potential field) 
are shown over the ten self-consistency cycles. For each cycle 20 Grad-Rubin
iterations are used to construct the $P$ and $N$ solutions. The energy of the 
$P$ solution is indicated by plus signs and a dashed line, and that of 
the $N$ solution by diamonds and a solid line. }
\label{f9}
\end{figure}

The energy of the $P$ and $N$ solution at each self-consistency 
cycle may be used to monitor the convergence of the procedure. 
At the end of each cycle (after $20$ Grad-Rubin iterations 
for both $P$ and $N$ solutions), the total magnetic energy, 
$E$, for the solutions is calculated. Figure \ref{f9} shows  

the evolution of $E/E_0$ with cycle for the $P$ and $N$ solutions. 
Initially the two energies are different (as discussed in Section 
\ref{res_24} and shown in Table \ref{t2}) but after three cycles the energy 
approaches a single value. After ten cycles $E/E_0=1.0047$ for 
the $P$ solution and $E/E_0=1.0044$ for the $N$ solution
which is lower than the energy of either the initial $P$ or $N$ solutions.
The small difference between the energies of the $P$ and $N$ solutions  
translates into a difference of $\approx 7\%$ in the free energy. In 
dimensional units the free energy is about $4\times 10^{29}\,{\rm erg}$. The 
total energy of the final self-consistent solution $E$, the energy of the 
potential field $E_0$, and the free energy of the final self-consistent 
solution $E_{\rm f}=E-E_0$ are listed in Table \ref{t3} (the first two rows 
of this table are relevant; the second two rows are discussed below). 

\begin{table}
\caption{Energy estimates for the self-consistent solutions constructed from the SOLIS/VSM
magnetogram data for AR 11029 on 24 October.}
\begin{tabular}{l  l  l  l  l }
\hline
\hline
Grad-Rubin iterations/cycle & Solution & $E$                      & $E_0$                   & $E_{\rm f}=E-E_0$  \\  
 &  & ($10^{31}\,{\rm erg}$ ) & ($10^{31}\,{\rm erg}$) & ($10^{29}\,{\rm erg}$) \\
\hline
$20$ & $P$  & $9.669$ & $9.62$ &  $4.52 $   \\
      & $N$  & $9.667$ & $9.62$ &  $4.23 $  \\

$35$ & $P$  & $9.666$ & $9.62$ &  $4.20 $   \\
      & $N$  & $9.667$ & $9.62$ &  $4.23 $  \\
\end{tabular}
\label{t3}
\end{table}

The lower panel of Figure \ref{f8} shows a visualization of the magnetic 
field of the self-consistent solutions. The magnetic field lines 
of the $P$ and $N$ solutions after ten self-consistency cycles are 
shown by the blue and red curves respectively. In contrast to the 
top panel of Figure \ref{f8}, the two 
sets of field lines are very similar. Small differences are seen for 
field lines extending high into the computational volume. The agreement between 
the two sets of field lines confirms that a single self-consistent 
solution is obtained.   

The oscillations of the energies of the initial $P$ and $N$ solutions 
(see Section \ref{res_24}) means that the magnetic field used to re-map $\alpha_0$ at the end of the 
early self-consistency cycle depends somewhat on the number of Grad-Rubin 
iterations used to construct the $P$ and $N$ solutions. If a 
different number of Grad-Rubin iterations is chosen the $P$ and $N$ solutions  
have different field lines, because of the oscillations seen in Figure \ref{f7}. 
Hence the mapping of field lines is different and the application 
of Equation \eq{alpha_new} gives a different result. Correspondingly 
the final energy of the self-consistency procedure depends somewhat  
on the number of Grad-Rubin iterations used. To test this dependence we 
repeat the calculation with $35$ Grad-Rubin iterations for the $P$ and 
$N$ solutions, instead of 20. We choose this value because it corresponds 
to an iteration at which the difference in energy between the $P$ and $N$ solutions
constructed directly from the vector magnetogram data is large (see Figure \ref{f7}). 
Once again ten self-consistency cycles are applied. The final energy and free energy in this case are also 
given in Table \ref{t3} in the third and fourth rows. The variation in the 
energies for the two self-consistent solutions with different numbers of 
Grad-Rubin iterations per self-consistency cycle is sufficiently small  
to justify an order of magnitude (or better) estimate for the free energy of 
the region based on the results. 

\begin{figure}
\centerline{\includegraphics[scale=0.85]{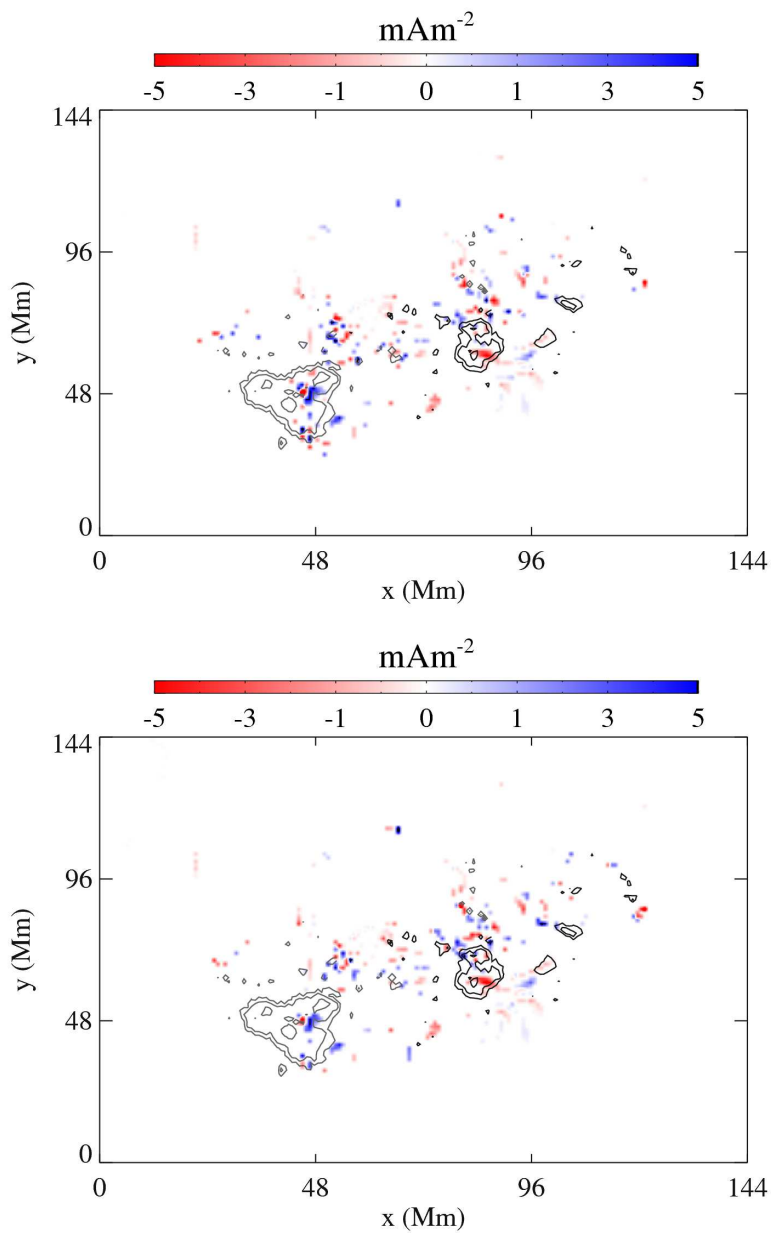}}
\caption{Results for the application of the self-consistency procedure
to the SOLIS/VSM vector magnetogram data for 24 October. The
vertical electric current density $J_z$ over part of the boundary is 
shown for the $P$ solution (upper panel),
and for the $N$ solution (lower panel) after ten self-consistency cycles.  
The values of $J_z$ are saturated at $\pm 5\,{\rm mAm^{-2}}$,
and only points where $|\alpha_0/\sigma|$ is three standard deviations above
the mean are shown. The contours are the same as Figure \ref{f2}. 
The distributions of $J_z$ for the $P$ and $N$ solutions after the 
self-consistency procedure are very similar, indicating a single 
self-consistent solution is found. }
\label{f10}
\end{figure}

The self-consistency procedure modifies $\alpha_0$ according to the 
procedure explained in Section \ref{intro_sc}. Figure \ref{f10} shows 
the vertical current density $J_z$ after ten self-consistency cycles for 
the $P$ solution (upper panel) and for the $N$ solution (lower panel). 
The self-consistent values of $J_z$ for the $P$ and $N$ solutions are very 
similar, confirming that a self-consistent force-free solution is 
found. This figure should be compared with the lower panel of Figure \ref{f2},
which shows $J_z$ for the original magnetogram data. Both Figures \ref{f2} and
\ref{f10} cover the same region of the photosphere.

The currents in the lower boundary are significantly reduced as a 
result of the self-consistency procedure (the maximum currents are $\approx 5\,{\rm mAm^{-2}}$
compared with $\approx 15\, {\rm mAm^{-2}}$ in the original data). This is 
in part due to the averaging implied by Equation \eq{alpha_new}. However,
it also represents the preservation of smaller but more certain values
and the loss of larger but less certain values (based on the phenomenological
uncertainty rule $\sigma \sim |B_z|^{-1}$). Specifically the values of $\alpha_0$ are 
typically smaller in strong field regions, which have a smaller uncertainty,
so these smaller values are better preserved in the application of 
Equation \eq{alpha_new} (compared to large values of $\alpha_0$ in weak 
field regions). There are substantial areas with $\alpha_0= 0$ in the final 
self-consistent boundary data. These patches correspond to 
open field regions where $\alpha$ is set to zero by the handling of 
the side and top boundary conditions, as explained in Section \ref{intro_forcefree}. 

\subsection{Construction of Self-consistent Solutions for 27 October}
\label{res_27}

\noindent The self-consistency procedure is applied to the {\it Hinode}/SP magnetogram data for 27 October. 
The initial $P$ and $N$ solutions again oscillate in energy. 
However, for the {\it Hinode}/SP data the oscillations are so severe 
that a meaningful estimate of the energy and free energy cannot be made. 
The large oscillations are due to large boundary values of $\alpha_0$. 
Many of these large values have a large associated uncertainty, and hence may be 
spurious, but the results of the calculation are influenced by these 
values. It is necessary to apply the self-consistency procedure 
which accounts for uncertainties in $\alpha_0$ and tends to remove values 
with large uncertainties in a systematic fashion. 

\begin{figure}
\centerline{\includegraphics[width=12cm,height=8cm]{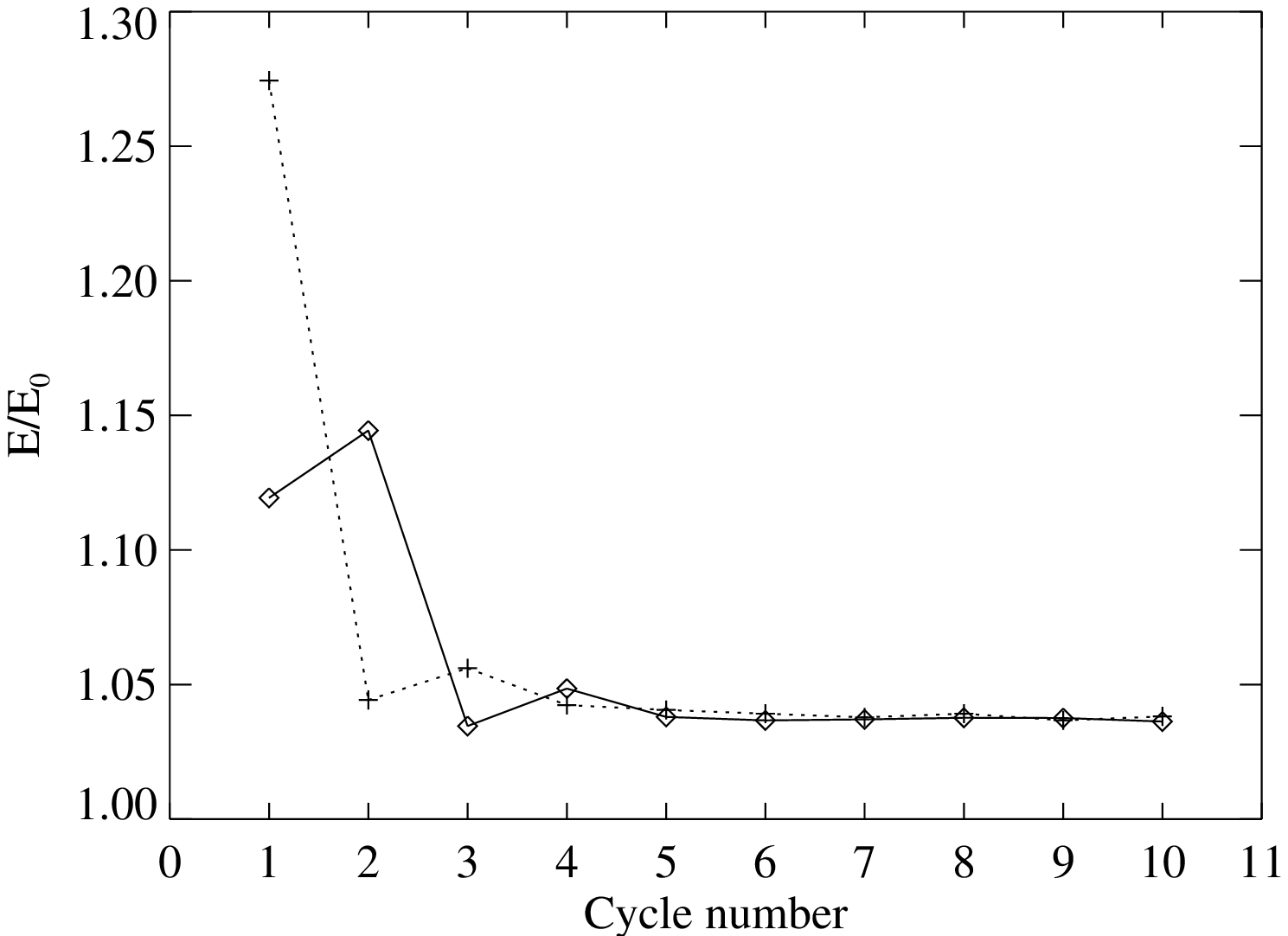}}
\caption{Results for the application of the self-consistency procedure to the
{\it Hinode}/SP vector magnetogram data for 27 October. The energies of 
the $P$ and $N$ solutions (as a fraction of the energy of the potential field) 
are shown over the ten self-consistency cycles. For each cycle 20 Grad-Rubin
iterations are used to construct the $P$ and $N$ solutions. The energy of the 
$P$ solution is indicated by plus signs and a dashed line, and that of 
the $N$ solution by diamonds and a solid line.}
\label{f11}
\end{figure}

The self-consistency procedure is applied to the {\it Hinode}/SP data for 27 October
using two separate runs with a different number of Grad-Rubin iterations per 
cycle for each run, similar to the procedure in Section \ref{sc_24} 
with the SOLIS/VSM data. For the first run we use $20$ Grad-Rubin iterations 
per self-consistency cycle, and for the second run we use $30$ Grad-Rubin 
iterations per self-consistency cycle. We apply ten self-consistency cycles in each case. 
For both runs the final energies of the $P$ and $N$ solution converge to a single value after
ten self-consistency cycles. Figure \ref{f11} shows the energy of the $P$ solution (plus signs) 
and $N$ solution (diamonds) in units of the potential field energy at the end of each 
self-consistency cycle over the ten cycles with 20 Grad-Rubin iterations 
per cycle. The results using 30 Grad-Rubin iterations are similar to those
using 20 iterations per cycle. In both cases the energies of the $P$ and $N$ solution initially differ 
greatly, but after ten self-consistency cycles the energy difference is 
reduced to $\approx 1\%$. 

The total energy, free energy of the $P$ and $N$ solutions, and the energy 
of the potential field for the self-consistency calculation are summarized 
in Table \ref{t4}. 

The total and free energy estimates differ for the two runs with different
numbers of Grad-Rubin iterations, but the 
difference is sufficiently small to allow an order of magnitude estimate for the energy. 
The total energy is large (of order $10^{33} \, 
{\rm erg}$, which is ten times larger than was found for the region on 24 October). 
The variation in the free energy is correspondingly large, but we can still 
place an upper bound of $\approx 10^{32} \, {\rm erg}$ on the free energy 
of the active region based on the results in Table \ref{t4}. 

\begin{table}[!h]
\caption{Energy estimates for the self-consistent solutions constructed from the {\it Hinode}/SP
magnetogram data for AR 11029 on 27 October.}
\begin{tabular}{l  l  l  l  l }
\hline
\hline
Grad-Rubin iterations/cycle & Solution & $E$                      & $E_0$                   & $E_{\rm f}=E-E_0$  \\  
 &  & ($10^{33}\,{\rm erg}$ ) & ($10^{33}\,{\rm erg}$) & ($10^{31}\,{\rm erg}$) \\
\hline
$20$ & $P$  & $1.769$ & $1.707$ &  $6.16 $   \\
      & $N$  & $1.772$ & $1.707$ &  $6.50 $  \\

$30$ & $P$  & $1.787$ & $1.707$ &  $7.94 $   \\
      & $N$  & $1.791$ & $1.707$ &  $8.35 $  \\

\end{tabular}
\label{t4}
\end{table}

\begin{figure}
\centerline{\includegraphics[width=12cm,height=10cm]{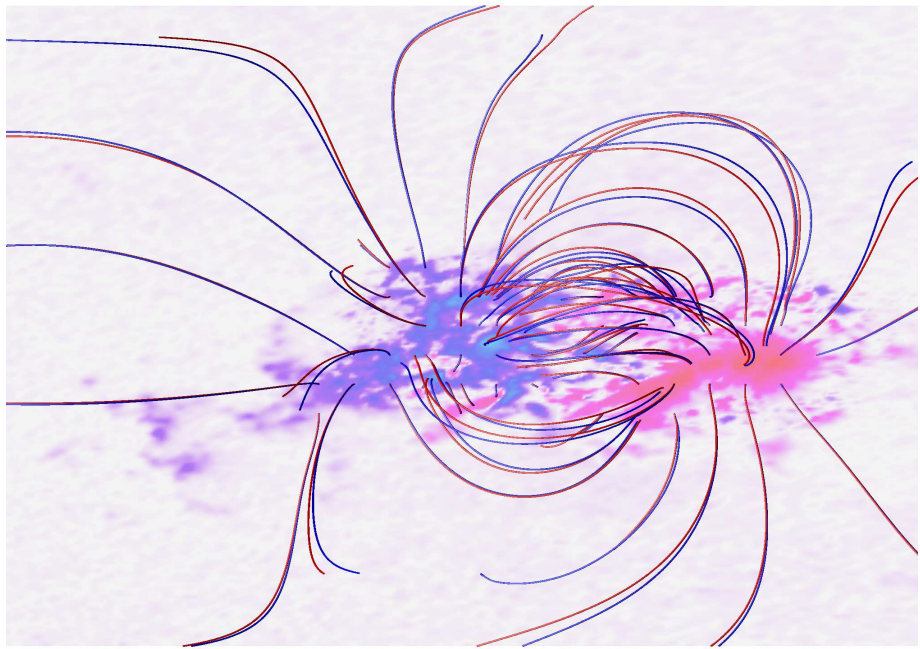}}
\caption{Nonlinear force-free models of the coronal magnetic field 
constructed from the {\it Hinode}/SP vector magnetogram data for 27 October. 
The self-consistent $P$ solution (blue field lines) and $N$ 
solution (red field lines) are shown for the calculation using ten 
self-consistency cycles with 20 Grad-Rubin iterations per cycle. The region is 
view looking down on the photosphere which is colored blue where $B_z>0$
and red where $B_z<0$.}
\label{f12}
\end{figure}

Figure \ref{f12} shows a visualization of the magnetic field of the 
self-consistent $P$ solution (blue curves) and $N$ solution (red curves) 
for the run using $20$ Grad-Rubin iterations per cycle. The two sets of field lines are similar, 
although some small differences are visible.

\begin{figure}
\centerline{\includegraphics[scale=0.85]{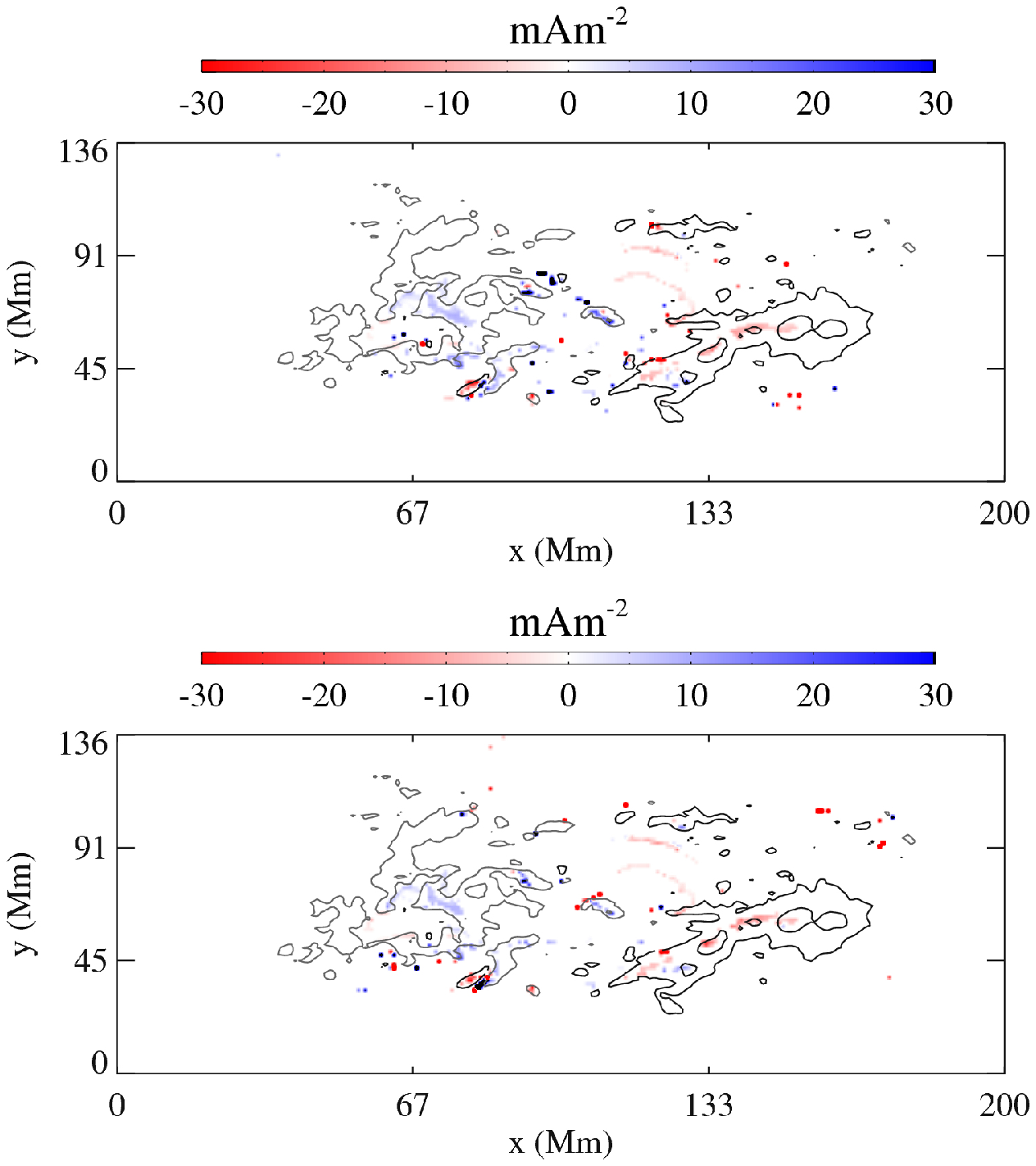}}
\caption{Results for the application of the self-consistency procedure
to the {\it Hinode}/SP vector magnetogram data for 27 October. The
vertical electric current density $J_z$ over part of the boundary is 
shown for the $P$ solution (upper panel),
and for the $N$ solution (lower panel) after ten self-consistency cycles.  
The values of $J_z$ are saturated at $\pm 30\,{\rm mAm^{-2}}$,
and only points where $|\alpha_0/\sigma|$ is three standard deviations above
the mean are shown. The contours are the same as Figure \ref{f4}. The distributions of $J_z$ for 
the $P$ and $N$ solutions after the self-consistency procedure are similar, 
indicating a single self-consistent solution is found. }
\label{f13}
\end{figure}

The self-consistency procedure modifies $\alpha_0$ according to the 
procedure explained in Section \ref{intro_sc}. Figure \ref{f13} shows 
the vertical current density $J_z$ after ten self-consistency cycles 
for the $P$ solution (upper panel) and for the $N$ solution (lower panel). 
The results are shown for the run using 20 Grad-Rubin iterations per cycle, but the results using $30$ 
Grad-Rubin iterations per cycle are similar. The self-consistent values 
of $J_z$ for the $P$ and $N$ solutions are very similar, confirming 
that a self-consistent solution is found. The self-consistent values of $J_z$
are significantly reduced by comparison with the values of $J_z$ in the
original magnetogram (lower panel of Figure \ref{f4}). This is due to the
averaging implicit in the procedure, and the removal by the procedure of large
values with large associated uncertainties (see discussion in Section \ref{intro_sc}).
The final self-consistent boundary data contains regions with $\alpha_0=0$ 
corresponding to open field regions where $\alpha$ has 
been set to zero by the handling of the side and top boundary conditions
(see Section \ref{intro_forcefree}).   

%
%
\section{Summary and Discussion of Results}
\label{sum_dis}

\noindent Estimates of the total energy and free energy 
of the coronal magnetic field of active region AR 11029 are calculated
using two vector magnetograms. The first magnetogram is based on data for 24 
October from the SOLIS Vector-SpectroMagnetograph (SOLIS/VSM), and the
second is based on data for 27 October obtained from the {\it Hinode} 
Solar Optical Telescope SpectroPolarimeter ({\it Hinode}/SP). The estimates 
use nonlinear force-free solutions constructed 
from the vector magnetogram boundary data. The magnetogram data provide two solutions to the nonlinear
force-free model (the $P$ and $N$ solutions, denoting the choice of polarity in the boundary 
conditions used). The model does not have a unique solution because 
the boundary data are inconsistent with the model. In each case
`self-consistent' solutions \citep{2009ApJ...700L..88W} are
constructed taking into account uncertainties in the boundary data. For 
the 24 October data the uncertainties are estimated phenomenologically, 
and for the 27 October data the uncertainties are obtained from the 
inversion process used to construct the magnetogram.  

The initial $P$ and $N$ solutions calculated from the vector 
magnetogram data for 24 October have different 
energies and different field line structures. These solutions are themselves
not uniquely defined because the Grad-Rubin iteration procedure used to 
construct the solutions does not strictly converge. The energy of the 
solutions oscillate with successive Grad-Rubin iterations owing to large
currents in the boundary data. The amplitude of the 
oscillations in the total energy is relatively small, but the
free energy is a small fraction of the total energy so this represents 
large amplitude oscillations in the free energy. The oscillations limit the 
accuracy of the estimation of the free energy, but an order of magnitude 
estimate is possible. The free energy of AR 11029 on 24 October is estimated 
from the initial $P$ and $N$ solutions to be $\approx 3 \times 10^{30}\,{\rm erg}$. 

The self-consistency procedure is applied to the 24 October data. The energy
of the $P$ and $N$ solutions agrees closely after ten self-consistency cycles,
and visual inspection of the field lines of the 
$P$ and $N$ solutions confirms that a self-consistent solution is obtained. 
The final energy of the self-consistent solution 
depends only weakly on the number of Grad-Rubin iterations 
per self-consistency cycle allowing a reliable estimate of the free 
energy. The free energy based on the self-consistent force-free solutions for 24 October is
$\approx 4\times 10^{29}\, {\rm erg}$. This value is roughly an order of 
magnitude smaller than the value obtained from the initial $P$ and $N$ 
solutions.  

For the 27 October {\it Hinode}/SP data, the initial $P$ and $N$ solutions do 
not permit a meaningful estimate of the free energy, because of large amplitude oscillations 
in the energy during the Grad-Rubin procedure. The misbehavior is 
attributed to large currents in the boundary data, and many of these 
values also have large associated uncertainties. The large values 
prevent convergence of the Grad-Rubin iteration, and hence 
the result is completely determined by the spurious boundary data. We 
believe the self-consistency procedure which accounts for these large 
uncertainties provides a more reliable result. The self-consistency procedure is applied 
to the {\it Hinode}/SP data leading to the estimate that the free energy of the 
region is $\approx 7\times10^{31} \, {\rm erg}$ on 27 October. 

The results presented here illustrate the practical difficulties with 
force-free modeling \citep{2009ApJ...696.1780D}. The models
constructed directly from the vector magnetogram data do not provide
a unique energy estimate so we use the self-consistency procedure.
The accuracy of the energy estimates obtained using the self-consistency 
procedure has not been tested in any independent 
way, and it is possible to identify aspects of the modeling which may lead
to underestimation of the energy. A number of factors tend to 
reduce the currents in the computational volume and in turn tend to reduce the free energy. 
Missing boundary data on the top and sides of the computational volume is handled by 
setting $\alpha=0$ along field lines which leave the top and sides of the volume.
This step removes currents from the volume. The self-consistency procedure 
also tends to reduce the $\alpha_0$ values on the lower boundary. This is 
due to three reasons. First, large values of $\alpha_0$ which have a large 
uncertainty are averaged out by the weighted average in the self-consistency
procedure. Second, regions with $\alpha_0 \ne 0$ connected by field lines 
to regions in the flux buffer with $\alpha_0 =0$ tend to get averaged to $\alpha_0=0$ over 
successive self-consistency cycles because points in the flux buffer
are assigned small uncertainties. Third, the boundary data have many 
small regions of $\alpha_0$ with opposite sign and averaging between 
these regions will lead to a net decrease in $\alpha_0$ over successive 
self-consistency cycles.

There is also considerable uncertainty associated with the boundary
magnetogram data. The field values are based on inversion
of spectropolarimetric measurements with finite spatial,
depth, and temporal resolution, and the inversions in this case assume a
Milne-Eddington atmosphere and the analytic form of the Stokes spectra
described by Unno and Rachkofsky ({\it e.g.} \citealt{2004ASSL..307.....L}).
The active region considered here, NOAA AR 11029,
evolved rapidly and was highly dynamic, producing Stokes spectra
which were strongly Doppler shifted and multi-lobed over much of
the region. The latter indicates that the observed spectra are not
consistent with a Milne-Eddington atmosphere, hence the returned field
values may not be appropriate estimates of
the underlying field.  The inversion based uncertainties (explained in
Section \ref{modeling_data2} ) for the {\it Hinode}/SP data reflect this
somewhat but are likely underestimates.  The inversions for the SOLIS/VSM
data can be assumed to harbor the same concerns. There is also 
uncertainty associated with the resolution of the 180-degree ambiguity 
\citep{2006SoPh..237..267M, 2009SoPh..260...83L}. 

\begin{figure}
\centerline{\includegraphics[width=10cm,height=8cm]{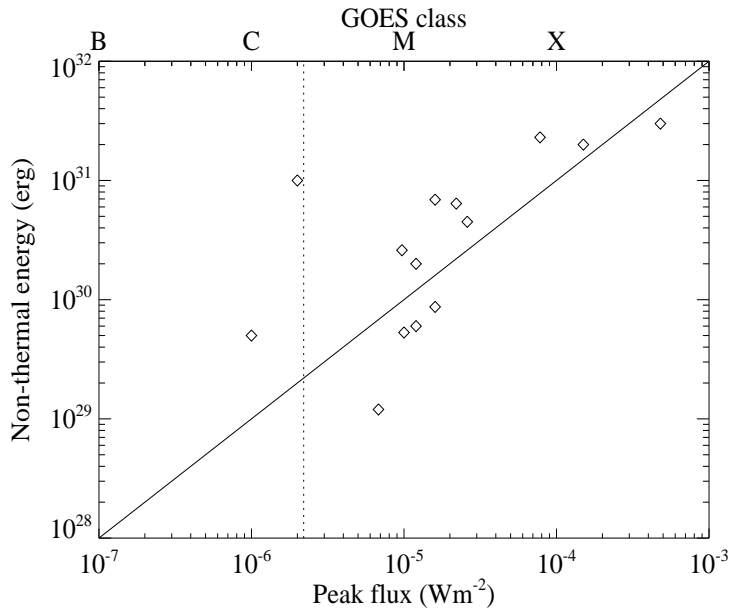}}
\caption{The relationship between RHESSI energy estimates for the non-thermal 
energy of a sample of flares from the literature (see Section \ref{sum_dis}) and the {\it GOES} 
peak flux of the flares. The plot also indicates the {\it GOES} classes. The vertical dashed
line shows the peak flux of a {\it GOES} C2.2 class flare (peak flux of $2.2\times 10^{-6}\,{\rm Wm^{-2}}$), 
the largest soft X-ray flare produced by AR 11029. }
\label{f14}
\end{figure}

Given these caveats, it is nevertheless interesting to compare the energy of the largest flares produced 
by AR 11029 with the free energy estimates derived for the region, although only a 
fraction of the free energy is expected to be released in any single flare.
To make this comparison we first establish an approximate correlation 
between {\it GOES} soft X-ray peak flux and flare energy. 
One measure of flare energy is the energy of non-thermal electrons, which 
may be inferred from hard X-ray observations \citep{1971SoPh...17..412L}.
We consider a sample from the literature of flare energies estimated from 
hard X-ray observations, restricting attention (for uniformity) to 
observations made by the Reuven Ramaty High-Energy Solar Spectroscopic 
Imager (RHESSI) \citep{2002SoPh..210....3L}. These flare energies may be 
compared with the observed {\it GOES} peak flux of each event to establish an
energy-peak flux scaling. Figure \ref{f14} shows the flare energy estimates made by 
\citet{2002SoPh..210..287S}, \citet{2005A&A...435..743S}, 
\citet{2006SoPh..235..125M}, \citet{2005AdSpR..35.1669H} and 
\citet{2004JGRA..10910104E} for 14 different RHESSI hard X-ray flares, 
plotted against the observed {\it GOES} soft X-ray peak flux of each flare. 
The plot shows that, to within an order of  magnitude, a {\it GOES} B-class 
event has an energy of $10^{28}\,{\rm erg}$, a C-class event has 
energy an energy of $10^{29}\,{\rm erg}$, an M-class event has an 
energy of $10^{30}\,{\rm erg}$, and an X-class event has energy above 
$10^{31}\,{\rm erg}$. This provides a useful rule-of-thumb scaling law for
flare energy estimates from {\it GOES} classification. These energy estimates
only take into account the energy of non-thermal electrons producing hard
X-rays in the flare. If an event is eruptive then a significant amount 
of energy can be carried off by the mass ejection \citep{2009EM&P..104..295G}, 
and active region AR 11029 produced many small eruptions during its lifetime\footnote{See the
LASCO CME catalog \url{http://cdaw.gsfc.nasa.gov/CME_list/}}.

On 24 October AR 11029 produced four B class 
flares, the largest of which is a B4.9 class event with a peak soft X-ray 
flux of $4.9\times 10^{-7}\,{\rm Wm^{-2}}$. The energy of each of these 
flares is estimated to be of the order $10^{28} \, {\rm erg}$, which is 
consistent with the upper bound of $4\times10^{29}\, {\rm erg}$ derived from 
the self-consistent force-free modeling and the upper bound of $3 \times 10^{30}\,{\rm erg}$ 
obtained from the initial $P$ and $N$ solutions (Tables \ref{t2} and \ref{t3}). 

On 27 October AR 11029 produced 23 soft X-ray flares. The largest
flare produced by AR 11029 was a {\it GOES} class C2.2 which occurred
at $00\colon24\,{\rm UT}$ on 28 October. The estimate
from the modeling for the free energy is $\approx 7\times 10^{31}\, {\rm erg}$ 
(Table \ref{t4}), which is comparable to an X-class flare.

Our estimates of the free energy of active region AR 11029 indicate that 
the region had sufficient energy to produce an M-class or X-class flare 
on 27 October, but not on 24 October. It is interesting that the 
estimate for the energy on 27 October is substantially larger than
that for 24 October. The photospheric extent of
the region grew rapidly during the interval 24 October to 27 October as
seen in Figure \ref{f0} and \citet{2010ApJ...710.1324W},
and it appears that the coronal magnetic field increased in size 
and energy accordingly. The magnetic flux of the region increased
significantly between 24 and 27 October. On 24 October the magnetic
flux was $\approx 5\times 10^{21}\,{\rm Mx}$ over the positive polarity and 
$\approx4\times 10^{21}\,{\rm Mx}$ over the negative polarity. On 27 October
the flux is $\approx30\times 10^{21}\,{\rm Mx}$ over both polarities, an
increase of a factor of 6-7. 

\section{Conclusion}

\noindent Solar active region AR 11029 was a highly flare productive 
sunspot region which emerged on the Sun in late October 2009. The region 
produced $73$ soft X-ray flares as recorded in the event list
for the {\it Geostationary Operational Environmental Satellites} ({\it GOES}) over a period of about a week 
(24 October to 1 November). Statistical analysis of these flares showed 
evidence for departure from the standard power-law frequency-size distribution
\citep{2010ApJ...710.1324W}. Specifically, an absence of large events was 
indicated. This was conjectured to be due to the small active region 
having insufficient magnetic energy to power large flares \citep{2010ApJ...710.1324W}.
The aim of this paper is to test this hypothesis by estimating the free 
energy ({\it i.e.} the energy available for flaring) using nonlinear force-free 
modeling of the coronal magnetic field of the active region from vector 
magnetogram data.  

Estimates of the magnetic free energy of the coronal magnetic field of 
AR 11029 are presented using vector magnetogram data taken by the  
Vector-SpectroMagnetograph at the Solar Observatory's Synoptic Long term 
Investigations of the Sun facility (SOLIS/VSM), and the 
Solar Optical Telescope SpectroPolarimeter on the {\it Hinode} 
satellite ({\it Hinode}/SP). Our data set consists of a SOLIS/VSM magnetogram taken on
24 October 2009 and a {\it Hinode}/SP magnetogram taken on 27 October 2009.
Force-free solutions are constructed from the data using the Grad-Rubin 
code due to \citet{2007SoPh..245..251W} and the self-consistency procedure
due to \citet{2009ApJ...700L..88W}. The procedure address the problem
of the inconsistency between the force-free model and the 
photospheric data (\citealt{1995ApJ...439..474M,2009ApJ...696.1780D}).

For the two vector magnetograms free energy estimates are obtained.
For 24 October we estimate a free energy of $\approx 4\times 10^{29}\, {\rm erg}$,
which is roughly the size of a {\it GOES} C-class or small {\it GOES} M-class flare. For
27 October we estimate a free energy $\approx 7\times 10 ^{31}\, {\rm erg}$, which
is roughly the size of a {\it GOES} X-class flare. Although free-energy
estimates made using non-linear force-free modeling are subject to significant
uncertainty we believe that the order of magnitude of these estimates is correct. 

The results in this paper do not support the hypothesis that 
there is an observable upper limit on the sizes of flares produced by 
small active regions due to insufficient free energy 
in the active region magnetic field. The modeling re-opens the interesting 
question of why this region did not produce any large flares given that it 
had sufficient energy to do so.  


\begin{acks}
S. A. Gilchrist acknowledges the support of an Australian Postgraduate Research Award. 
KDL appreciates funding from NSF SHINE grant ATM-0454610 and line-of-sight
potential field code from Graham Barnes. The authors thank 
Prof. Don Melrose for a thorough reading of the manuscript.
SOLIS data used here are produced cooperatively by NSF/NSO and NASA/LWS. 
{\it Hinode} is a Japanese mission developed 
and launched by ISAS/JAXA, collaborating with NAOJ as a domestic partner, 
and NASA and STFC (UK) as international partners. Scientific operation of 
the {\it Hinode} mission is conducted by the {\it {\it Hinode}} science team 
organized at ISAS/JAXA. This team mainly consists of scientists from 
institutes in the partner countries. Support for the post-launch operation 
is provided by JAXA and NAOJ (Japan), STFC (U.K.), NASA, ESA, and 
NSC (Norway). 
\end{acks}


\appendix   

\noindent Glossary of terms.

\begin{itemize}
\item Model: the nonlinear force-free model.
\item Solution: the magnetic field $\mathbf B = \mathbf B(\mathbf r)$ and the 
  force-free parameter $\alpha=\alpha(\mathbf r)$ (where $\mathbf r$ 
  is the position vector) which solve Equations \eq{ff_lor} and 
  \eq{ff_ampere} in a domain with boundary conditions for $B_z$ and $\alpha$ (over a single polarity
  of $B_z$) prescribed on the boundary of the domain. 
\item $P$ Solution: a solution where boundary conditions
for $\alpha$ are prescribed at boundary points where 
$B_z>0$. 
\item $N$ Solution: a solution where boundary conditions
for $\alpha$ are prescribed at boundary points where 
$B_z<0$. 
\item Iteration: one step in a procedure, {\it e.g.} one Grad-Rubin iteration is
  going from $k$ to $k+1$ in the Grad-Rubin or current-field iteration
  procedure, as defined in \citet{2007SoPh..245..251W}.
\item Self-consistency cycle: construction of $P$ and $N$ Grad-Rubin solutions
  (involving a certain number of Grad-Rubin iterations in each case) from given
  $B_z$ and $\alpha_0$ boundary conditions (see \citealt{2009ApJ...700L..88W}).
\item Fixed point/convergence: a fixed point in the Grad-Rubin procedure is
  defined by $\mathbf B^{[k]}= \mathbf B^{[k+1]}$ for iterations $k$ and $k+1$ 
  (see \citealt{2007SoPh..245..251W}). This represents convergence of the procedure
  (when met for all points in a computational volume).
\end{itemize}


\end{article} 
\end{document}